\documentclass[10pt,conference,twocolumn]{IEEEtran}

\usepackage{amsmath}
\usepackage{amssymb}
\usepackage{latexsym}
\usepackage{amsfonts}

\begin{document}

\pagestyle{headings}  
\vspace{1cm}

%
\title{The $4$-error linear complexity distribution for $2^n$-periodic binary
sequences}

\author{
\authorblockN{Jianqin Zhou}
\authorblockA{ Department of Computing, Curtin University, Perth, WA 6102 Australia\\
 Computer Science School, Anhui Univ. of
Technology, Ma'anshan, 243002 China\\ \ \ zhou9@yahoo.com\\
\ \\
Jun Liu\\
Telecommunication School, Hangzhou Dianzi University,
Hangzhou, 310018 China\\
\ \\
Wanquan Liu\\
Department of Computing, Curtin University, Perth, WA 6102 Australia\\
 w.liu@curtin.edu.au
             }
        }
\maketitle              

\begin{abstract}
By using the sieve method of combinatorics, we study $k$-error linear complexity distribution of $2^n$-periodic binary sequences based on Games-Chan algorithm.  For $k=4,5$, the complete counting functions on
the $k$-error linear complexity of $2^n$-periodic balanced binary
sequences (with linear complexity less than $2^n$) are
presented.
As a
consequence of the result,
the complete counting functions on
the $4$-error linear complexity of $2^n$-periodic binary
sequences (with linear complexity $2^n$ or less than $2^n$) are obvious.
Generally, the complete counting functions on the $k$-error linear complexity of $2^n$-periodic binary sequences can be obtained with a similar approach.


\end{abstract}

\section{Introduction}

The linear complexity of a sequence $s$, denoted as
$L(s)$, is defined as the length of the
shortest linear feedback shift register (LFSR) that can generate $s$. The concept of linear complexity is very useful in the
study of security of stream ciphers for cryptographic applications
and it has attracted much attention in cryptographic community
\cite{Ding,Stamp}.
Ding, Xiao and Shan  first in their book
\cite{Ding} proposed the weight complexity and sphere
complexity. Stamp and Martin \cite{Stamp} introduced $k$-error
linear complexity, which is similar to the sphere complexity, and
put forward the concept of $k$-error linear complexity profile.
Specifically, suppose that $s$ is a sequence  with period $N$. For
any $k(0\le k\le N)$, $k$-error linear complexity of $s$, denoted as
$L_k(s)$,  is defined as the smallest linear complexity that can be
obtained when any $k$ or fewer bits of the sequence are changed
within one period.

One important result, proved by Kurosawa et al. in \cite{Kurosawa}
is that the minimum number $k$ for which the $k$-error linear
complexity of a $2^n$-periodic binary sequence $s$ is strictly less
than a linear complexity $L(s)$ of $s$ is determined by
$k_{\min}=2^{W_H(2^n-L(s))}$, where $W_H(a)$ denotes the Hamming
weight of the binary representation of an integer $a$.
In \cite{Meidl}, for the period length $p^n$, where $p$ is an odd prime and 2 is a primitive root modulo $p^2$, the  relationship is showed between the linear complexity and the minimum value $k$ for which the $k$-error linear complexity is strictly less than the linear complexity.
In \cite{Zhou},
for sequences over $GF(q)$ with period $2p^n$, where $p$ and $q$ are odd primes, and $q$ is a primitive root modulo $p^2$, the minimum value $k$ is presented for which the $k$-error linear complexity is strictly less than the linear complexity.

Rueppel
\cite{Rueppel} derived  the number $N(L)$ of $2^n$-periodic binary
sequences with given linear complexity $L, 0\le L \le 2^n$.
For $k=1,2$, Meidl \cite{Meidl2005} characterized the complete
counting functions on the $k$-error linear complexity of
$2^n$-periodic binary sequences having maximal possible linear
complexity $2^n$. For $k=2,3$, Zhu and Qi \cite{Zhu} further showed
the complete counting functions on the $k$-error linear complexity
of $2^n$-periodic binary sequences with linear complexity $2^n-1$.
By using algebraic and combinatorial methods, Fu et al. \cite{Fu}
studied the linear complexity and the $1$-error linear complexity
for $2^n$-periodic binary sequences, and characterized the
 $2^n$-periodic binary sequences with given 1-error linear
 complexity and derived the counting function for the 1-error
 linear complexity for $2^n$-periodic binary sequences.

By investigating sequences with linear complexity $2^n$ or linear
complexity less than $2^n$ together, Kavuluru
\cite{Kavuluru2008,Kavuluru} characterized $2^n$-periodic binary
sequences with fixed 2-error or 3-error linear complexity,  and
obtained the counting functions for the number of $2^n$-periodic
binary sequences with given $k$-error linear complexity for $k = 2$
and 3.  In \cite{Zhou2,Zhou_Liu}, it is  proved that the counting functions in
\cite{Kavuluru2008,Kavuluru} for the number of $2^n$-periodic binary
sequences with  3-error linear complexity are incorrect in some
cases.
The complete counting functions for the number of $2^n$-periodic binary
sequences with  3-error linear complexity  are
presented in \cite{Zhou_Liu}.

In this paper, based on the sieve method of combinatorics, we will study the $k$-error linear complexity by
 an approach different from those in current literature
\cite{Meidl2005,Zhu,Kavuluru2008,Kavuluru}. We investigate sequences
with  linear complexity $2^n$,  and sequences with linear complexity
less than $2^n$, separately.  It is observed that for sequences with
linear complexity $2^n$, the $k$-error linear complexity is equal to
$(k+1)$-error linear complexity, when $k$ is odd. For sequences with
linear complexity less than $2^n$, the $k$-error linear complexity
is equal to  $(k+1)$-error linear complexity, when $k$ is even.

Therefore, to characterize $2^n$-periodic binary  sequences
with fixed 4-error linear complexity,  first we consider $2^n$-periodic binary  sequences  with
linear complexity $2^n$ and fixed 3-error linear complexity, which is the result of \cite{Zhou_Liu}, then in this paper we consider $2^n$-periodic binary  sequences  with
linear complexity less than $2^n$ and fixed 4-error linear complexity.

Let $S=\{s | L(s)=c\}, E=\{e | W_H(e)\le k\}, SE=\{s+e | s\in S,
e\in E\}$, where $s$ is a sequence with linear complexity $c$, and
$e$ is an error sequence with $W_H(e)\le k$. With the sieve method
of combinatorics, we sieve sequences $s+e$  with $L_k(s+e)=c$ from
$SE$.

 For $k=4,5$, the complete counting functions on
the $k$-error linear complexity of $2^n$-periodic binary
sequences (with linear complexity less than $2^n$) are
presented.
By combining the results in \cite{Zhou_Liu}: the complete counting functions on
the $3$-error linear complexity of $2^n$-periodic binary
sequences (with linear complexity  $2^n$),
the complete counting functions on
the $4$-error linear complexity of $2^n$-periodic binary
sequences (with linear complexity $2^n$ or less than $2^n$) are obvious.

The rest of this paper is organized as follows. In  Section II, some
preliminary results are presented.
 In Section III, for $k=4,5$, the  counting functions on the $k$-error linear
complexity of $2^n$-periodic binary sequences (with linear
complexity less than $2^n$) are  characterized.

\section{Preliminaries}

In this section we give some preliminary results which will be used in the sequel.

We will consider sequences over $GF(q)$, which is the finite field
of order $q$. Let $x=(x_1,x_2,\cdots,x_n)$ and
$y=(y_1,y_2,\cdots,y_n)$ be vectors over $GF(q)$. Then define
$x+y=(x_1+y_1,x_2+y_2,\cdots,x_n+y_n)$.

When $n=2m$, we define $Left(x)=(x_1,x_2,\cdots,x_m)$ and
$Right(x)=(x_{m+1},x_{m+2},\cdots,x_{2m})$.

The Hamming weight of an $N$-periodic sequence $s$ is defined as the
number of nonzero elements in per period of $s$, denoted by
$W_H(s)$. Let $s^N$ be one period of $s$. If $N=2^n$, $s^N$ is also
denoted as $s^{(n)}$. Obviously, $W_H(s^{(n)})=W_H(s^N)=W_H(s)$.
The distance of two elements is defined as the difference of their indexes.

The generating function of a sequence $s=\{s_0, s_1, s_2, s_3,
\cdots, \}$  is defined by $$s(x)=s_0+ s_1x+ s_2x^2+ s_3x^3+
\cdots=\sum\limits^\infty_{i=0}s_ix^i$$

The generating function of a finite sequence $s^N=\{s_0, s_1, s_2,
 \cdots, s_{N-1},\}$ is defined by $s^N(x)=s_0+ s_1x+ s_2x^2+
\cdots+s_{N-1}x^{N-1}$. If $s$ is a periodic sequence with the first
period $s^N$, then,
\begin{eqnarray}
s(x) &=& s^N(x)(1+ x^N+ x^{2N}+ \cdots)=\frac{s^N(x)}{1-x^N}\notag\\
&=&\frac{s^N(x)/\gcd(s^N(x),1-x^N)}{(1-x^N)/\gcd(s^N(x),1-x^N)}\notag\\
&=&\frac{g(x)}{f_s(x)}\label{formula01}
\end{eqnarray}
where $f_s(x)=(1-x^N)/\gcd(s^N(x),1-x^N),
g(x)=s^N(x)/\gcd(s^N(x),1-x^N)$.

Obviously, $\gcd(g(x),f_s(x))=1, \deg(g(x))<\deg(f_s(x))$. $f_s(x)$
is called  the minimal polynomial of $s$, and the degree of $f_s(x)$
is called the linear complexity of $s$, that is $\deg(f_s(x))=L(s)$.

Suppose that $N=2^n$ and $GF(q)=GF(2)$. Then
$1-x^N=1-x^{2^n}=(1-x)^{2^n}=(1-x)^N$. Thus for binary sequences
with period $2^n$, to compute linear complexity is  to find the degree of
factor $(1-x)$ in $s^N(x)$.

The following three lemmas are  well known results on $2^n$-periodic
binary sequences.

\noindent {\bf Lemma  2.1} Suppose that $s$ is a binary sequence with
period $N=2^n$. Then $L(s)=N$ if and only if the Hamming weight of a
period of the sequence is odd.

If an element one is removed from a sequence whose Hamming weight is
odd, the Hamming weight of the sequence will be changed to even, so
the main concern hereinafter is about sequences whose Hamming weight
are even.

\noindent {\bf Lemma 2.2}  Let $s_1$ and $s_2$ be two binary sequences
with period $N=2^n$. If $L(s_1)\ne L(s_2)$, then
$L(s_1+s_2)=\max\{L(s_1),L(s_2)\} $; otherwise if $L(s_1)= L(s_2)$,
then $L(s_1+s_2)<L(s_1)$.

Suppose that the linear complexity of $s$ can decline when at least
$k$ elements of $s$ are changed. By Lemma 2.2, the linear complexity
of the binary sequence, in which elements at exactly those $k$
positions are all nonzero, must be $L(s)$. Therefore, for the
computation of $k$-error linear complexity, we only need to find the
binary sequence whose Hamming weight is minimum and its linear
complexity is $L(s)$.

\noindent {\bf Lemma  2.3} Let $E_i$ be a $2^n$-periodic sequence
with one nonzero element at position $i$ and 0 elsewhere in each
period, $0\le i<2^n$. If $j-i=2^r(1+2a), a\ge0, 0\le i<j<2^n,
r\ge0$, then $L(E_i +E_j)=2^n-2^r$.

\noindent {\bf Lemma  2.4} Suppose that $s$ is a binary sequence
with period $N=2^n$ and the Hamming weight is $4$. Then the linear
complexity of $s$ is $L(s)=2^n-2^{n-m}, 1<m\le n$ or
$2^n-(2^{n-m}+2^{n-j})$, $1\le m<j\le n$.

\begin{proof}\
Suppose that a $2^n$-periodic binary sequence $s$ with $W_H(s)=4$.
If the  linear complexity of $s$ is neither $2^n-2^{n-m}$ nor
$2^n-(2^{n-m}+2^{n-j})$, then the minimum number $k$ for which the
$k$-error linear complexity of a $2^n$-periodic binary sequence $s$
is strictly less than  linear complexity $L(s)$  can be given by
$k_{\min}=2^{W_H(2^n-L(s))}>4$, where $W_H(b)$ denotes the Hamming
weight of the binary representation of an integer $b$,  which
contradicts the fact that the $4$-error linear complexity of the
binary sequence $s$ is $0$ ($LC_4(s)=0$). So the linear complexity
of $s$ is $2^n-2^{n-m}, 1<m\le n$ or $2^n-(2^{n-m}+2^{n-j})$,
$1\le m<j\le n$.

\end{proof}\

\

\section{Counting functions with
the $4$-error linear complexity}

In \cite{Zhou_Liu}, the $3$-error linear complexity of $2^n$-periodic binary
sequences with  linear complexity  $2^n$ is discussed. For $2^n$-periodic binary sequences with  linear complexity  $2^n$, the change of 4 bits per period results in a
sequence with odd number of nonzero bits per period, which has again
linear complexity $2^n$. Therefore, the  $4$-error linear complexity of $2^n$-periodic binary
sequences with  linear complexity  $2^n$ is the same as the $3$-error linear complexity.
In order to derive the  counting functions of the $4$-error linear complexity for $2^n$-periodic binary
sequences (with linear complexity $2^n$ or less than $2^n$), we  only need to investigate the $4$-error linear complexity of $2^n$-periodic binary
sequences with  linear complexity less than $2^n$.

The main result of this section is the following theorem.

 \noindent {\bf Theorem  3.1}  Let $L(r,c)=2^n-2^r+c$, $2\le r\le
n, 1\le c\le 2^{r-1}-1$, and $N_4(L)$ be the number of
$2^n$-periodic binary sequences with  linear complexity less than
$2^n$ and given 4-error linear complexity $L$. Then

{\scriptsize
$
N_4(L)=\left\{\begin{array}{l}
1+\left(\begin{array}{c}2^{n}\\2\end{array}\right)+\left(\begin{array}{c}2^{n}\\4\end{array}\right),\\
\ \ \ \ \ \ \ \ L=0  \\
2^{L(r,c)-1}(1+\left(\begin{array}{c}2^{r}\\2\end{array}\right)+\left(\begin{array}{c}2^{r}\\4\end{array}\right)),\\
\ \ \ \ \ \ \ \ L=L(r,c), 1\le c\le 2^{r-3}-1, r>3\\
2^{L(r,c)-1}f(r,m),  \\
\ \ \ \ \ \ \ \ L=L(r,c),  c= 2^{r-2}-2^{r-m},2<m\le r, r>2\\
2^{L(r,c)-1}g(r,m),  \\
\ \ \ \ \ \ \ \ L=L(r,c),  c= 2^{r-2}-2^{r-m}+x,\\
 \ \ \ \ \  \ \ \ \ \  2<m<r-1,0<x<2^{r-m-1},r>4\\
2^{L(r,c)-1}h(r,m),  \\
\ \ \ \ \ \ \ \ L=L(r,c),  c= 2^{r-1}-2^{r-m},2\le m\le r, r\ge 2\\
2^{L(r,c)-1}p(r,m,j),\\
\ \ \ \ \ \ \ \ \  L=L(r,c),  c= 2^{r-1}-(2^{r-m}+2^{r-j}),\\
 \ \ \ \ \  \ \ \    2<m<j\le r, r>3\\
2^{L(r,c)-1}q(r,m,j),\\
\ \ \ \ \ \ \ \ \  L=L(r,c),  c= 2^{r-1}-(2^{r-m}+2^{r-j})+x,\\
 \ \ \ \ \  \ \ \    2<m<j<r-1,0<x<2^{r-j-1},r>5\\
0, \ \ \ \ \ \ \ \ \ \mbox{otherwise}
\end{array}\right.
$} where $f(r,m)$, $g(r,m)$, $h(r,m)$, $p(r,m,j)$, $q(r,m,j)$ are
defined in the following Lemma 3.5, 3.6, 3.7, 3.9 and 3.10 respectively.

\

To prove Theorem  3.1, we need to give several lemmas.

Given a $2^n$-periodic binary sequence $s$, its linear complexity $L(s)$
can be  determined by the Games-Chan algorithm \cite{Games}. Based
on Games-Chan algorithm, the following  Lemma  3.1 is given in
\cite{Meidl2005}.

\noindent {\bf Lemma  3.1} Suppose that $s$ is a binary sequence with
first period $s^{(n)}=\{s_0,s_1,s_2,\cdots, s_{2^n-1}\}$, a mapping
$\varphi_n$ from $F^{2^n}_2$ to $F^{2^{n-1}}_2$ is given by
\begin{eqnarray*}&&\varphi_n(s^{(n)})\\
&=&\varphi_n((s_0,s_1,s_2,\cdots,
s_{2^n-1}))\\
&=&(s_0+s_{2^{n-1}},s_1+s_{2^{n-1}+1},\cdots,
s_{2^{n-1}-1}+s_{2^n-1}) \end{eqnarray*}

Let $W_H(\mathbf{\upsilon})$ denote the Hamming weight of a vector
$\mathbf{\upsilon}$. Then mapping $\varphi_n$ has the following
properties

1) $W_H(\varphi_n(s^{(n)}))\le W_H(s^{(n)})$;

2) If $n\ge2$, then $W_H(\varphi_n(s^{(n)}))$ and $W_H(s^{(n)})$ are
either both odd or both even;

3) The set $$\varphi^{-1}_{n+1}(s^{(n)})=\{v\in
F^{2^{n+1}}_2|\varphi_{n+1}(v)=s^{(n)} \}$$ of the preimage of
$s^{(n)}$ has cardinality $2^{2^n}$.

Rueppel \cite{Rueppel} presented the following.

\noindent {\bf Lemma  3.2}  The number $N(L)$ of $2^n$-periodic
binary sequences with given linear complexity $L, 0\le L \le 2^n$,
is given by $N(L)=\left\{\begin{array}{l}
1, \ \ \ \ \ L=0\ \   \\
2^{L-1}, \ 1\le L\le 2^n
\end{array}\right.$\

\

It is known that the computation of $k$-error linear complexity can be converted
to finding error sequences with minimal Hamming weight. Hence  4-error linear complexity of $s^{(n)}$ is the smallest
linear complexity that can be obtained when any $u^{(n)}$ with
$W_H(u^{(n)})=0$, 2 or 4 is added to $s^{(n)}$.
 Let $s^{(n)}$ be a binary sequence
with linear complexity $c$, $u^{(n)}$ is a binary sequence with $W_H(u^{(n)})\le k$.
We derive the counting functions on the $k$-error linear complexity of $2^n$-periodic binary
sequences by investigating $s^{(n)}+u^{(n)}$.
Based on this idea, we first prove the following lemmas.

\noindent {\bf Lemma  3.3} Suppose that $s^{(n)}$ and $t^{(n)}$ are
two different binary sequences with linear complexity $c, 1\le
c\le 2^{n-3}$, and   $u^{(n)}$ and $v^{(n)}$  are two different  binary
sequences, $W_H(u^{(n)})=0$, 2 or 4, $W_H(v^{(n)})=0$, 2 or 4.
Then $s^{(n)}+u^{(n)}\ne t^{(n)}+v^{(n)}$.

\begin{proof}\ The following is obvious

$s^{(n)}+u^{(n)}\ne t^{(n)}+v^{(n)}$

$\Leftrightarrow$ $s^{(n)}+u^{(n)}+v^{(n)}\ne t^{(n)}$

$\Leftrightarrow$ $u^{(n)}+v^{(n)}\ne s^{(n)}+t^{(n)}$

Note that $s^{(n)}$ and $t^{(n)}$ are two different binary
sequences with linear complexity $c, 1\le c\le 2^{n-3}$, so the
linear complexity of $s^{(n)}+t^{(n)}$ is less than $2^{n-3}$, hence
one period of $s^{(n)}+t^{(n)}$ can be divided into 8 equal parts.

Suppose that $u^{(n)}+v^{(n)}= s^{(n)}+t^{(n)}$. Then one period of
$u^{(n)}+v^{(n)}$ can be divided into 8 equal parts, thus it has 8
nonzero elements. It follows that the linear complexity of
$u^{(n)}+v^{(n)}$ is $2^{n-3}$, which contradicts the fact that the
linear complexity of $s^{(n)}+t^{(n)}$ is less than $2^{n-3}$.
\end{proof}\

\noindent {\bf Lemma  3.4}  1). If $s^{(n)}$ is a binary sequence
with linear complexity $c, 1\le c\le 2^{n-1}-3$, $c\ne
2^{n-1}-2^{n-m}, 1<m<n-1$ and $c\ne 2^{n-1}-(2^{n-m}+2^{n-j})$,
$1<m<j\le n$, $u^{(n)}$ is a  binary sequence, and $W_H(u^{(n)})=0$,
2 or 4. Then the 4-error linear complexity of $s^{(n)}+u^{(n)}$ is
still $c$.

2). If $s^{(n)}$ is a binary sequence with linear complexity
$c=2^{n-1}-2^{n-m}, 1<m\le n$ or $c=2^{n-1}-(2^{n-m}+2^{n-j})$,
$1<m<j\le n$, then there exists a binary sequence $u^{(n)}$ with
$W_H(u^{(n)})=0,2$ or 4, such that the 4-error linear complexity of
$s^{(n)}+u^{(n)}$ is less than $c$.

\begin{proof}\
Without loose of  generality, we suppose that  $v^{(n)}\ne u^{(n)}$, and  $W_H(v^{(n)})=0$, 2 or 4.

1). As $1\le c\le 2^{n-1}-3$, we only need to consider the case
$L(u^{(n)}+v^{(n)})<2^{n-1}$. Thus
$Left(u^{(n)}+v^{(n)})=Right(u^{(n)}+v^{(n)})$ and
$W_H(Left(u^{(n)}+v^{(n)}))=2$ or 4.

 By Lemma 2.3 and Lemma 2.4, $L(u^{(n)}+v^{(n)})=2^{n-1}-2^{n-m}$, or $2^{n-1}-(2^{n-m}+2^{n-j})$.

Thus $L(s^{(n)}+u^{(n)}+v^{(n)})\ge L(s^{(n)})$, so the 4-error
linear complexity of $s^{(n)}+u^{(n)}$ is  $c$.

2). As $s^{(n)}$ is a binary sequence with  linear complexity
$c=2^{n-1}-2^{n-m},  1<m\le n$ or $c=2^{n-1}-(2^{n-m}+2^{n-j})$,
$1<m<j\le n$. So the 4-error linear complexity of $s^{(n)}+u^{(n)}$
must be less than $c$ when $L(u^{(n)}+v^{(n)})=c$.
\end{proof}\

Next we divide the 4-error linear complexity into six categories and deal with them respectively.
First consider the category of $2^{n-2}-2^{n-m}$.

\noindent {\bf Lemma  3.5}  Let  $N_4(2^{n-2}-2^{n-m})$ be the
number of $2^n$-periodic binary sequences with linear complexity
less than $2^n$ and given 4-error linear complexity
$2^{n-2}-2^{n-m}, n>2, 2<m\le n$. Then{\scriptsize 
\begin{eqnarray*}&&N_4(2^{n-2}-2^{n-m})\\
&=&\left[\begin{array}{l}1+\left(\begin{array}{c}2^n\\2\end{array}\right)+\left(\begin{array}{c}2^n\\4\end{array}\right)-C1-C2/2\end{array}\right]
\times2^{2^{n-2}-2^{n-m}-1}\end{eqnarray*}} where C1, C2 are defined
in  the following proof.

\begin{proof}\

We first  sketch  the proof.
Let $S=\{s | L(s)=2^{n-2}-2^{n-m}\}, E=\{e | W_H(e)=0 \mbox{ or } 2\mbox{ or } 4\}, S+E=\{s+e | s\in S, e\in E\}$, where $s$ is a sequence with
linear complexity $2^{n-2}-2^{n-m}$, and $e$ is an error sequence with $W_H(e)=0$ or $2$ or $4$. With the sieve method of combinatorics, we attempt to sieve sequences $s+e$
with $L_4(s+e)=2^{n-2}-2^{n-m}$ from $S+E$.

If $W_H(e)=1 \mbox{ or } 3$, then $W_H(s+e)$ is odd, thus $L(s+e)=2^n$. As we consider binary sequences with linear complexity
less than $2^n$, so we only consider error sequences with $W_H(e)=0$ or $2$ or $4$.

By Lemma  3.2, the number  of $2^n$-periodic binary sequences with linear complexity $2^{n-2}-2^{n-m}$ is
$2^{2^{n-2}-2^{n-m}-1}$. As the number  of $2^n$-periodic binary sequences in $E$ is $1+\left(\begin{array}{c}2^n\\2\end{array}\right)+\left(\begin{array}{c}2^n\\4\end{array}\right)$,
 the  number  of $2^n$-periodic binary sequences $s+e\in S+E$
is at most $(1+\left(\begin{array}{c}2^n\\2\end{array}\right)+\left(\begin{array}{c}2^n\\4\end{array}\right)) 2^{2^{n-2}-2^{n-m}-1}$.

It remains to characterize two cases. One is that $s+e\in S+E$, but $L_4(s+e) <2^{n-2}-2^{n-m}$. The other  is the case that
 $L_4(s^{(n)}+u^{(n)})=L_4(t^{(n)}+v^{(n)})=2^{n-2}-2^{n-m}$ with $s^{(n)}\ne t^{(n)}$, $u^{(n)}\ne v^{(n)}$, but $s^{(n)}+u^{(n)}= t^{(n)}+v^{(n)}$.

The following is the detailed proof.

Obviously, by Lemma  3.4, there exists 2 distinct binary sequences $u^{(n)}$ and $v^{(n)}$  with
 $W_H(u^{(n)})=W_H(v^{(n)})=4$, such that
$L(u^{(n)}+v^{(n)})=2^{n-2}-2^{n-m}$.
Then the 4-error linear complexity of $u^{(n)}+s^{(n)}$ can be less than $2^{n-2}-2^{n-m}$.

First, suppose that $w^{(n)}$ is a  binary sequence with linear
complexity $2^{n-2}-2^{n-m}$ and $W_H(w^{(n)})=8$. Then $w^{(n)}$ can be divided into 4 equal parts and the number
of these sequences is given by

\begin{eqnarray*}
&&(2^{n-2}-2^{n-m})+(2^{n-2}-2^{n-m}\times3)\\
&&\ \ \ \ \ \ \ +\cdots+(2^{n-2}-2^{n-m}\times(\frac{2^{n-2}}{2^{n-m}}-1))\\
&=&\frac{2^{n-2}}{2}\times\frac{2^{n-2}}{2\times2^{n-m}}\\
&=&2^{n+m-6}
\end{eqnarray*}

 The following example is given to illustrate the above formula.

 Suppose that $n=5,m=4$,  and $L(w^{(n)})=2^{n-2}-2^{n-m}$. Then the first $2^{n-2}=8$ elements of $w^{(n)}$ can  be

 $\{1010\ 0000\}$,  $\{0101\ 0000\}$, $\{0010\ 1000\}$, $\{0001\ 0100\}$,

$\{0000\ 1010\}$, $\{0000\ 0101\}$,  $\{1000\ 0010\}$, $\{0100\ 0001\}$.

Each $w^{(n)}$ has the intersection of 4 nonzero elements with $2(\frac{2^{n-2}}{2\times2^{n-m}}-1)=2(2^{m-3}-1)$
other different $w^{(n)}$.  All these 4-nonzero-element constitute  the
set

\begin{eqnarray*}&&A_1=\{(a_i,a_{i+2^{n-2}},a_{i+2^{n-1}},a_{i+2^{n-1}+2^{n-2}})|\\
&&\ \ \ \ \ \ \ \ \ 0\le
i<2^{n-2}\}\end{eqnarray*}

So the number of elements in  $A_1$ is $2^{n-2}$.

If  the 4 nonzero elements of $u^{(n)}$ are part of one of these
$w^{(n)}$, where $W_H(w^{(n)})=8$, then the number of these
$u^{(n)}$ can be given by
$$C1=2^{n+m-6}\left(\begin{array}{c}8\\4\end{array}\right)-2^{n-2}(2^{m-3}-1)$$

Note that the 4 nonzero elements of $u^{(n)}$ can  be in  $A_1$.\\

Second, suppose that $w^{(n)}$ is a binary sequence with linear
complexity $2^{n-2}-2^{n-k}$, $2<k<m$, and $W_H(w^{(n)})=8$. We
define $A_2$ as the set of these sequences.  So the number of these
sequences is given by $|A_2|=2^{n+k-6}$.

If  the 4 nonzero bits of $u^{(n)}$ are part of $w^{(n)}\in A_2$,
but does not belong to $A_1$. Then the number of these $u^{(n)}$ can
be given by
$2^{n+k-6}[\left(\begin{array}{c}8\\4\end{array}\right)-2].$

For $2<k<m$, the total number of $u^{(n)}$ can be given by {\small
$C2=\sum\limits_{k=3}^{m-1}
2^{n+k-6}(\left(\begin{array}{c}8\\4\end{array}\right)-2)=(2^{n+m-6}-2^{n-3})(\left(\begin{array}{c}8\\4\end{array}\right)-2)$}

Of these $u^{(n)}$, there exists exactly one distinct binary sequence $v^{(n)}$ with $W_H(v^{(n)})=4$, such that $L(u^{(n)}+v^{(n)})=2^{n-2}-2^{n-k}$,$2<k<m$.
Let $t^{(n)}=s^{(n)}+u^{(n)}+v^{(n)}$.
Then $L(t^{(n)})=L(s^{(n)})=2^{n-2}-2^{n-m}$ and $t^{(n)}+v^{(n)}=s^{(n)}+u^{(n)}$.

This leads to the following

{\scriptsize 
\begin{eqnarray*}&&N_4(2^{n-2}-2^{n-m})\\
&=&[1+\left(\begin{array}{c}2^n\\2\end{array}\right)+\left(\begin{array}{c}2^n\\4\end{array}\right)-C1-C2/2]\times2^{2^{n-2}-2^{n-m}-1}\end{eqnarray*}}
\end{proof}\

We define $N_4(2^{n-2}-2^{n-m})=f(n,m)\times2^{2^{n-2}-2^{n-m}-1}$\\

Next we  consider  the category of $2^{n-2}-2^{n-m}+x$.

 \noindent {\bf Lemma
3.6}  Let $N_4(2^{n-2}-2^{n-m}+x)$ be the number of $2^n$-periodic
binary sequences with linear complexity less than $2^n$ and given
4-error linear complexity $2^{n-2}-2^{n-m}+x, n>4,2<m<n-1,
0<x<2^{n-m-1}$. Then{\small
\begin{eqnarray*}&&N_4(2^{n-2}-2^{n-m}+x)\\
&=&[1+\left(\begin{array}{c}2^n\\2\end{array}\right)+\left(\begin{array}{c}2^n\\4\end{array}\right)-\frac{2^{m-2}-1}{2^{m-2}}D1
-\frac{1}{2}D2]\\
&&\ \ \times2^{2^{n-2}-2^{n-m}+x-1}\end{eqnarray*}} where D1, D2 are
defined in  the following proof.

\begin{proof}\
Suppose that $s^{(n)}$ is a binary sequence with linear complexity
$$2^{n-2}-2^{n-m}+x=2^{n-1}-(2^{n-2}+2^{n-m})+x, $$ and $u^{(n)}$ is a binary sequence with
$W_H(u^{(n)})=0, 2$ or 4. By Lemma 3.4, the 4-error linear
complexity of $s^{(n)}+u^{(n)}$ is still $2^{n-2}-2^{n-m}+x$.

By Lemma 3.4, there exist 2 distinct binary sequences $u^{(n)}$ and
$v^{(n)}$, with $W_H(u^{(n)})=W_H(v^{(n)})=4$, such that
$L(u^{(n)}+v^{(n)})<2^{n-2}-2^{n-m}+x$. Let
$t^{(n)}=s^{(n)}+u^{(n)}+v^{(n)}$. Then
$L(t^{(n)})=L(s^{(n)})=2^{n-2}-2^{n-m}+x$ and
$t^{(n)}+v^{(n)}=s^{(n)}+u^{(n)}$.

Suppose that $w^{(n)}$ is a binary sequence with linear complexity
$2^{n-2}-2^{n-k}$, $2<k\le m$ and $W_H(w^{(n)})=8$.
Then the number of these sequences is given by $2^{n+k-6}$.
Each $w^{(n)}$ has the intersection of 4 nonzero elements with
$2(\frac{2^{n-2}}{2\times2^{n-k}}-1)=2(2^{k-3}-1)$
other different ones.  All these 4-nonzero-element constitute  the
set
\begin{eqnarray*}&&A_3=\{(a_i,a_{i+2^{n-2}},a_{i+2^{n-1}},a_{i+2^{n-1}+2^{n-2}})|\\
&&\ \ \ \ \ \ \ \ \ 0\le
i<2^{n-2}\}\end{eqnarray*}

So the number of elements in  $A_3$ is $$D1=2^{n-2}$$

Suppose that $u^{(n)}$ is a binary sequence with $W_H(u^{(n)})=4$ and its 4 nonzero bits belong to $A_3$.
Then there exist $\sum\limits_{k=3}^m \frac{2^{n-2}}{2^{n-k+1}}=\sum\limits_{k=3}^m 2^{k-3}=2^{m-2}-1$
  distinct binary sequences $v^{(n)}$ with $W_H(v^{(n)})=4$,
such that $L(u^{(n)}+v^{(n)})=2^{n-2}-2^{n-k}$, $2<k\le m$.
Let $t^{(n)}=s^{(n)}+u^{(n)}+v^{(n)}$.
Then  $L(t^{(n)})=L(s^{(n)})=2^{n-2}-2^{n-m}+x$, and $s^{(n)}+u^{(n)}=t^{(n)}+v^{(n)}$.

On the other side, if the  4 nonzero elements of sequence $u^{(n)}$ are part of the 8 nonzero elements of a $w^{(n)}$
but does not belong to $A_3$, then there exists exactly one binary sequence $v^{(n)}$ with $W_H(v^{(n)})=4$,
such that $L(u^{(n)}+v^{(n)})=2^{n-2}-2^{n-k}$, $2<k\le m$.
The number of these $u^{(n)}$ can be given by
{\small
$D2=\sum\limits_{k=3}^{m}
2^{n+k-6}(\left(\begin{array}{c}8\\4\end{array}\right)-2)=(2^{n+m-5}-2^{n-3})(\left(\begin{array}{c}8\\4\end{array}\right)-2)$}

By Lemma 3.2, the number of $2^{n}$-periodic binary sequences with given linear complexity
$2^{n-2}-2^{n-m}+x$ is $2^{2^{n-2}-2^{n-m}+x-1}$.
This will derive the following
{\small
\begin{eqnarray*}&&N_4(2^{n-2}-2^{n-m}+x)\\
&=&[1+\left(\begin{array}{c}2^n\\2\end{array}\right)+\left(\begin{array}{c}2^n\\4\end{array}\right)-\frac{2^{m-2}-1}{2^{m-2}}D1 -\frac{1}{2} D2]\\
&&\ \ \times2^{2^{n-2}-2^{n-m}+x-1}\end{eqnarray*}}
\end{proof}\

We define {\small
$$N_4(2^{n-2}-2^{n-m}+x)=g(n,m)\times2^{2^{n-2}-2^{n-m}+x-1}$$}

We  consider the category of $2^{n-1}-2^{n-m}$.

\noindent {\bf Lemma  3.7}  Let $N_4(2^{n-1}-2^{n-m})$ be the number
of $2^n$-periodic binary sequences with linear complexity less than
$2^n$ and given 4-error linear complexity $2^{n-1}-2^{n-m},  2\le
m\le n$. Then {\scriptsize
\begin{eqnarray*}&&N_4(2^{n-1}-2^{n-m})\\
&=&[\left(\begin{array}{c}2^n\\4\end{array}\right)-E1+E2/4-E3+E4/2-E5+E6/4-E7+E8/8]\\
&&\ \ \times2^{2^{n-1}-2^{n-m}-1}\end{eqnarray*}} where $E1,
E2,\cdots,E8$ are defined in  the following proof.

\begin{proof}\
Suppose that $s^{(n)}$ is a binary sequence with linear complexity
$2^{n-1}-2^{n-m}$, and $u^{(n)}$ is a binary sequence with
$W_H(u^{(n)})=0$ or 2. It is easy to verify that there exists a binary sequence
$v^{(n)}$  with  $W_H(v^{(n)})=4$, such that
$L(u^{(n)}+v^{(n)})=2^{n-1}-2^{n-m}$, then the 4-error linear
complexity of $u^{(n)}+s^{(n)}$ is less than $2^{n-1}-2^{n-m}$.

Suppose that $u^{(n)}$ is a binary sequence with $W_H(u^{(n)})=4$.
Let us divide one period of $u^{(n)}$ into $2^{n-m}$ subsequences
of form
$$\{u_a,u_{a+2^{n-m}}, u_{a+2^{n-m+1}},\cdots,
u_{a+(2^m-1)\times2^{n-m}}\}$$ where $0\le a<2^{n-m}.$

 Case 1): If only 2 nonzero elements of $u^{(n)}$ are in one subsequence and
the other 2 nonzero elements are in another one, then the number of these $u^{(n)}$ can be given by
$$E1=\left(\begin{array}{c}2^{n-m}\\2\end{array}\right)\times\left(\begin{array}{c}2^{m}\\2\end{array}\right)
\times\left(\begin{array}{c}2^{m}\\2\end{array}\right)$$

Of these $u^{(n)}$, if the distance of  2 nonzero bits in each subsequence   is not $2^{n-m}(1+2a)$, $a\ge0$,
in other words, 2 nonzero bits in each subsequence are both at odd locations or even locations,
then the number of these $u^{(n)}$ can be given by

$\left(\begin{array}{c}2^{n-m}\\2\end{array}\right)\times\left(\begin{array}{c}2\\1\end{array}\right)\times\left(\begin{array}{c}2^{m-1}\\2\end{array}\right)
\times\left(\begin{array}{c}2\\1\end{array}\right)\times\left(\begin{array}{c}2^{m-1}\\2\end{array}\right)$.

There are
{\scriptsize
$$\left(\begin{array}{c}2^{n-m}\\2\end{array}\right)\times \left(\begin{array}{c}2\\1\end{array}\right)\times 2^{m-1} \times \left(\begin{array}{c}2\\1\end{array}\right)(\left(\begin{array}{c}2^{m-1}\\2\end{array}\right)-2^{m-2})$$}
sequences, of which there are exactly 2 nonzero elements in each sequence whose distance  is $2^{n-1}$.

Therefore, there exist
{\scriptsize
$$\left(\begin{array}{c}2^{n-m}\\2\end{array}\right)\times[2^{m-1}\times2^{m-1}
+2^{m+1}(\left(\begin{array}{c}2^{m-1}\\2\end{array}\right)-2^{m-2})]$$}
sequences, of which there are at least 2 nonzero elements in each sequence whose distance  is $2^{n-1}$.

So, if only 2 nonzero elements are in each subsequence,  and the distance of each pair of nonzero elements in each subsequence is neither $2^{n-m}(1+2a)$ nor $2^{n-1}$, then the number of these $u^{(n)}$ can be given by
\begin{eqnarray*}
&&E2=4\times\left(\begin{array}{c}2^{n-m}\\2\end{array}\right)\times\left(\begin{array}{c}2^{m-1}\\2\end{array}\right)\times\left(\begin{array}{c}2^{m-1}\\2\end{array}\right)\\
&& \ \ \ \ \ \ \ -\left(\begin{array}{c}2^{n-m}\\2\end{array}\right)\times[2^{2m-2}+2^{m+1}(\left(\begin{array}{c}2^{m-1}\\2\end{array}\right)\\
&& \ \ \ \ \ \ \ -2^{m-2})]
\end{eqnarray*}

There exist 3 distinct sequences $v^{(n)}_i, 1\le i\le3$, with $W_H(v^{(n)}_i)=4$, such that
$L(u^{(n)}+v^{(n)}_i)=2^{n-1}-2^{n-r}$ or $2^{n-1}-(2^{n-r}+2^{n-k})$, $1<r<m, m+1\le k\le n$.
Let $t^{(n)}_i=s^{(n)}+u^{(n)}+v^{(n)}_i$.
Then $L(t^{(n)}_i)=L(s^{(n)})=2^{n-1}-2^{n-m}$ and $s^{(n)}+u^{(n)}=t^{(n)}_i+v^{(n)}_i$.

To illustrate the construction of $v^{(n)}$, the following example is given.

 Suppose that $n=5,m=4$, and  $Left(u^{(n)})=\{1000\ 1001\ 0000\ 0001\}$. Then let $Right(v^{(n)})=Left(u^{(n)})$ and $Left(v^{(n)})=Right(u^{(n)})$, thus $L(u^{(n)}+v^{(n)})=2^{n-1}-2^{n-3}$.

$Left(u^{(n)}+v^{(n)})$ can also be $\{1000\ 1000\ 0000\ 0000\}$ or $\{0000\ 0001\ 0000\ 0001\}$.

 Case 2): If only 2 nonzero elements of $u^{(n)}$ are in one subsequence and
the other 2 nonzero elements are in 2 other distinct subsequences,
then the number of these $u^{(n)}$ can be given by
$$E3=\left(\begin{array}{c}2^{n-m}\\3\end{array}\right)\times\left(\begin{array}{c}3\\1\end{array}\right)\times\left(\begin{array}{c}2^{m}\\2\end{array}\right)\times2^m\times2^m$$

Of these $u^{(n)}$, the distance of the 2 nonzero bits which are in one  subsequence is not $2^{n-m}(1+2a)$,
then the number of these $u^{(n)}$ can be given by $\left(\begin{array}{c}2^{n-m}\\3\end{array}\right)\times\left(\begin{array}{c}3\\1\end{array}\right)\times(\left(\begin{array}{c}2^{m-1}\\2\end{array}\right)\times2)\times2^{m}\times2^{m}$.
And there are
$\left(\begin{array}{c}2^{n-m}\\3\end{array}\right)\times\left(\begin{array}{c}3\\1\end{array}\right)\times2^{m-1}\times2^m\times2^m$
sequences, in which the distance of the 2 nonzero bits  is $2^{n-1}$.

So, if only 2 nonzero elements are in one  subsequence, the other 2 nonzero elements are in 2 other distinct ones.
The distances of the 2 nonzero elements which are in one  subsequences is neither $2^{n-m}(1+2a)$ nor $2^{n-1}$,
then the number of these $u^{(n)}$ can be given by
\begin{eqnarray*}
&&E4=\left(\begin{array}{c}2^{n-m}\\3\end{array}\right)\left(\begin{array}{c}3\\1\end{array}\right)\left(\begin{array}{c}2^{m-1}\\2\end{array}\right)\times2^{2m+1} \\
&&\ \ \ \ \ \ \ \ -\left(\begin{array}{c}2^{n-m}\\3\end{array}\right)\left(\begin{array}{c}3\\1\end{array}\right)\times2^{3m-1}\end{eqnarray*}

There exist a sequence $v^{(n)}$, with $W_H(v^{(n)})=4$, such that
$L(u^{(n)}+v^{(n)})=2^{n-1}-2^{n-r}$, $1<r<m$.

 Case 3): If there are only 3 nonzero elements  in one subsequence,
then the number of these $u^{(n)}$ can be given by
$$ E5=\left(\begin{array}{c}2^{n-m}\\2\end{array}\right)\left(\begin{array}{c}2\\1\end{array}\right)\left(\begin{array}{c}2^m\\3\end{array}\right)\times2^m $$

Suppose that there are only 3 nonzero elements of $u^{(n)}$  in one  subsequence, and there do not exist 2 nonzero bits whose distance
is $2^{n-m}(1+2a)$. Then the number of these $u^{(n)}$ can be given by $\left(\begin{array}{c}2^{n-m}\\2\end{array}\right)\left(\begin{array}{c}2\\1\end{array}\right)[\left(\begin{array}{c}2^{m-1}\\3\end{array}\right)\times2]\times2^m$.

Of these $u^{(n)}$, there are
$$\left(\begin{array}{c}2^{n-m}\\2\end{array}\right)\left(\begin{array}{c}2\\1\end{array}\right)[2^{m-1}\times(2^{m-1}-2)]\times2^{m}$$ sequences,
in which there exist 2 nonzero elements whose distance is $2^{n-1}$.

So,if there are only 3 nonzero elements of $u^{(n)}$  in one subsequence, and there do not exist 2 nonzero elements
whose distance is $2^{n-m}(1+2a)$ or $2^{n-1}$, then the number of these $u^{(n)}$ can be given by \begin{eqnarray*}
&&E6=2^{m+2}\times\left(\begin{array}{c}2^{n-m}\\2\end{array}\right)\times\left(\begin{array}{c}2^{m-1}\\3\end{array}\right)\\
&&\ \ \ \ \ \ \ \ -\left(\begin{array}{c}2^{n-m}\\2\end{array}\right)\times(2^{m-1}-2)\times2^{2m}\end{eqnarray*}

There exist 3 distinct binary sequences $v^{(n)}_i, 1\le i\le3$, with $W_H(v^{(n)}_i)=4$, such that
$L(u^{(n)}+v^{(n)}_i)=2^{n-1}-2^{n-r}$ , $1<r<m$.
Let $t^{(n)}_i=s^{(n)}+u^{(n)}+v^{(n)}_i$.
Then $L(t^{(n)}_i)=L(s^{(n)})=2^{n-1}-2^{n-m}$ and $s^{(n)}+u^{(n)}=t^{(n)}_i+v^{(n)}_i$.

To illustrate the construction of $v^{(n)}$, the following example is given.

 Suppose that $n=5,m=4$, and  $Left(u^{(n)})=\{1000\ 1000\ 0000\ 1001\}$. Then
$Left(u^{(n)}+v^{(n)})$ can  be $\{1000\ 1000\ 0000\ 0000\}$, $\{1000\ 0000\ 0000\ 1000\}$  or $\{0000\ 1000\ 0000\ 1000\}$.

 Case 4): If all 4 nonzero elements of $u^{(n)}$ are in one subsequence, then the number of these $u^{(n)}$ can be given by $$E7=2^{n-m}\times\left(\begin{array}{c}2^{m}\\4\end{array}\right)$$

Suppose that all 4 nonzero elements of $u^{(n)}$ are in one subsequence, and there do not exist 2 nonzero elements whose distance
is $2^{n-m}(1+2a)$. Then the number of these $u^{(n)}$ can be given by
$2^{n-m}\times2\times\left(\begin{array}{c}2^{m-1}\\4\end{array}\right)$.

Of these $u^{(n)}$, there are
$$2^{n-m}\times2\times2^{m-2}[\left(\begin{array}{c}2^{m-1}-2\\2\end{array}\right)-2^{m-2}+1]
$$
sequences,
in which there exist exactly  2 nonzero elements with distance  $2^{n-1}$.

There are
$$2^{n-m+1}\times\left(\begin{array}{c}2^{m-2}\\2\end{array}\right)
$$
sequences,
in which there exist exactly  2 pair of elements with distance  $2^{n-1}$.

Thus there are
$$2^{n-1}\left(\begin{array}{c}2^{m-1}-2\\2\end{array}\right)
-2^{n-m+1}\times\left(\begin{array}{c}2^{m-2}\\2\end{array}\right)$$ sequences,
in which there exist at least 2 nonzero elements whose distance is $2^{n-1}$.

So, if all 4 nonzero elements of $u^{(n)}$ are in one subsequence, and there do not exist 2 nonzero elements
whose distance is $2^{n-m}(1+2a)$ or $2^{n-1}$, then the number of these $u^{(n)}$ can be given by \begin{eqnarray*}
&&E8=2^{n-m+1}\times\left(\begin{array}{c}2^{m-1}\\4\end{array}\right)-[2^{n-1}
 \times\left(\begin{array}{c}2^{m-1}-2\\2\end{array}\right)\\
 &&\ \ \ \ \ \ \ \ -2^{n-m+1}\times\left(\begin{array}{c}2^{m-2}\\2\end{array}\right)]
\end{eqnarray*}

There exist $\left(\begin{array}{c}4\\2\end{array}\right)+1=7$ distinct binary sequences $v^{(n)}_i, 1\le i\le7$, with $W_H(v^{(n)}_i)=4$, such that
$L(u^{(n)}+v^{(n)}_i)=2^{n-1}-2^{n-r}$, $1<r<m$ or $2^{n-1}-(2^{n-r}+2^{n-k})$, $1<r<k<m$.
Let $t^{(n)}_i=s^{(n)}+u^{(n)}+v^{(n)}_i$.
Then $L(t^{(n)}_i)=L(s^{(n)})=2^{n-1}-2^{n-m}$ and $s^{(n)}+u^{(n)}=t^{(n)}_i+v^{(n)}_i$.

To illustrate the construction of $v^{(n)}$, the following example is given.

 Suppose that $n=5,m=4$, and  $Left(u^{(n)})=\{1000\ 1000\ 1000\ 1000\}$. Then let $Right(v^{(n)})=Left(u^{(n)})$ and $Left(v^{(n)})=Right(u^{(n)})$, thus $L(u^{(n)}+v^{(n)})=2^{n-1}-(2^{n-2}+2^{n-3})$.

$Left(u^{(n)}+v^{(n)})$ can also be
$\{1000\ 1000\ 0000\ 0000\}$, $\{1000\ 0000\ 1000\ 0000\}$, $\{1000\ 0000\ 0000\ 1000\}$, $\{0000\ 1000\ 1000\ 0000\}$,$\{0000\ 1000\ 0000\ 1000\}$, $\{0000\ 0000\ 1000\ 1000\}$.

By Lemma  3.2,  the number  of $2^n$-periodic binary sequences with
given linear complexity $2^{n-1}-2^{n-m}$ is
$2^{2^{n-1}-2^{n-m}-1}$. It follows that,

{\scriptsize
\begin{eqnarray*}&&N_4(2^{n-1}-2^{n-m})\\
&=&[\left(\begin{array}{c}2^n\\4\end{array}\right)-E1+E2/4-E3+E4/2-E5+E6/4-E7+E8/8]\\
&&\ \  \times2^{2^{n-1}-2^{n-m}-1}\end{eqnarray*}}
\end{proof}\

We define $N_4(2^{n-1}-2^{n-m})=h(n,m)\times2^{2^{n-1}-2^{n-m}-1}$\\

Here we present an important lemma, which will be used in the
following discussions.

\noindent {\bf Lemma  3.8} Suppose that $s^{(n)}$ is a $2^n$-periodic binary sequence
with linear complexity $2^{n-1}-(2^{n-m}+2^{n-j})$, $n>3, 2<m<j\le n$, and $W_H(s^{(n)})=8$.
Then the number of these $s^{(n)}$ is $2^{n+2m+j-10}$.

\begin{proof}\
Suppose that $s^{(x)}$ is a $2^x$-periodic binary sequence with linear complexity  $2^x-2^y$,
 $y<x$, and $W_H(s^{(x)})=2$.
Then it is easy to verify that the number of these $s^{(x)}$ can be given by
$$(2^x-2^y)+[2^x-2^y(1+2)]+\cdots+2^y=2^{2x-y-2}$$
So the number of $2^{x+1}$-periodic binary sequences $s^{(x+1)}$ with linear complexity $2^{x+1}-(2^x+2^y)=2^x-2^y$ and $W_H(s^{(x+1)})=4$ also
can be given by $2^{2x-y-2}$.

For $n-1>x+1$, based on Games-Chan algorithm,
if $2^{n-1}$-periodic binary sequences $s^{(n-1)}$ with linear complexity $2^{n-1}-(2^x+2^y)$ and $W_H(s^{(n-1)})=4$,
then the number of these $s^{(n-1)}$ can be given by
$(2^4)^{n-x-2}\times2^{2x-y-2}=2^{4n-2x-y-10}$.

So, if $x=n-m, y=n-j$, then the number of $2^n$-periodic binary sequences $s^{(n)}$ with linear complexity $2^{n-1}-(2^x+2^y)$ and $W_H(s^{(n)})=8$ can be given by $$2^{4n-2x-y-10}=2^{n+2m+j-10}$$
\end{proof}\

We are ready to investigate the category of
$2^{n-1}-(2^{n-m}+2^{n-j})$.

\noindent {\bf Lemma  3.9}  Let $N_4(2^{n-1}-(2^{n-m}+2^{n-j}))$ be the number of $2^n$-periodic binary sequences
with linear complexity less than $2^n$ and given 4-error linear complexity $2^{n-1}-(2^{n-m}+2^{n-j}), n>3, 2<m<j\le n$.
Then
{\scriptsize
\begin{eqnarray*}&&N_4(2^{n-1}-(2^{n-m}+2^{n-j}))\\
&=&[1+\left(\begin{array}{c}2^n\\2\end{array}\right)+\left(\begin{array}{c}2^n\\4\end{array}\right)-F4\\
&&-\sum\limits_{k=m+1}^{j-1}(\frac{2^{2m-3}-1}{2^{2m-3}}F6+\frac{2^{m-1}-1}{2^{m-1}}F7+F8/2)\\
&&-\frac{2^{m-2}-1}{2^{m-2}}F10-F11/2-F13-\frac{3}{4}F14-\frac{2^{m-1}-1}{2^{m-1}}F17\\
&&-\frac{3}{4}F18-\frac{2^{2m-4}-1}{2^{2m-4}}F19-F22/2-\frac{2^{m-2}-1}{2^{m-2}}F23\\
&&-F25-F26-\frac{7}{8}F27]\times2^{2^{n-1}-(2^{n-m}+2^{n-j})-1}
\end{eqnarray*}}
where $F1, F2,\cdots,F27$ are defined in  the following proof.

\begin{proof}\
Suppose that $s^{(n)}$ is a binary sequence  with linear complexity $2^{n-1}-(2^{n-m}+2^{n-j})$,
$u^{(n)}$ is a binary sequence with $W_H(u^{(n)})=0$,2 or 4. Then the number of these $u^{(n)}$ can be given by
$$1+\left(\begin{array}{c}2^n\\2\end{array}\right)+\left(\begin{array}{c}2^n\\4\end{array}\right)$$

Next we will investigate the $u^{(n)}$ by Case A) and Case B).

Case A): Suppose that $w^{(n)}$ is a binary sequence with linear complexity $2^{n-1}-(2^{n-m}+2^{n-j})$, and $W_H(w^{(n)})=8$. By Lemma 3.8,
then the number of these $w^{(n)}$ can be given by $2^{n+2m+j-10}$.

By Lemma 3.4, there exist sequences $u^{(n)}$ and $v^{(n)}$ with $W_H(u^{(n)})=W_H(v^{(n)})=4$,
such that $L(u^{(n)}+v^{(n)})=2^{n-1}-(2^{n-m}+2^{n-j})$,
thus the 4-error linear complexity of $u^{(n)}+s^{(n)}$ is less than  $2^{n-1}-(2^{n-m}+2^{n-j})$.

Suppose that $u^{(n)}$ is a binary sequence with $W_H(u^{(n)})=4$,
and  the 4 nonzero bits are part of  the 8 nonzero elements of a $w^{(n)}$.
There are totally $2^{n+2m+j-10}\times\left(\begin{array}{c}8\\4\end{array}\right)$ such sequences $u^{(n)}$,
but some of which are counted repeatedly.

Case A.1):  $Left(u^{(n)})=Right(u^{(n)})$.
Of these $u^{(n)}$, if there exist 2 nonzero elements in $Left(u^{(n)})$ whose distance is $2^{n-j}(1+2a)$,
then the number of these $u^{(n)}$ can be given by $$F1=2^{n-j}\times\left(\begin{array}{c}2^{j-1}\\2\end{array}\right)-2^{n-j+1}\times\left(\begin{array}{c}2^{j-2}\\2\end{array}\right)=2^{n+j-4}$$

And each of these $u^{(n)}$ is counted $\frac{2^{n-1}}{2^{n-m+1}}\times\frac{2^{n-1}}{2^{n-m+1}}=2^{2(m-2)}$ times repeatedly in $2^{n+2m+j-10}\times\left(\begin{array}{c}8\\4\end{array}\right)$.

To illustrate the construction of $w^{(n)}$, the following example is given.

 Suppose that $n=5,m=3, j=4$, and  $Left(u^{(n)})=\{1010\ 0000\ 0000\ 0000\}$. Then
$Left(w^{(n)})$ can  be $\{1010\ 1010\ 0000\ 0000\}$, $\{1010\ 1000\ 0000\ 0010\}$, $\{1010\ 0010\ 0000\ 1000\}$ or $\{1010\ 0000\ 0000\ 1010\}$.

If the distance of 2 nonzero elements in $Left(u^{(n)})$ is $2^{n-m}(1+2a)$, then each of these $u^{(n)}$ is counted $2^{2(m-2)}\times\frac{2^{n-m}}{2^{n-j+1}}=2^{m+j-5}$ times repeatedly in $2^{n+2m+j-10}\times\left(\begin{array}{c}8\\4\end{array}\right)$,
and the number of these $u^{(n)}$ can be given by
\begin{eqnarray*}&&
F2=2^{n-m}\left(\begin{array}{c}2^{m-1}\\2\end{array}\right)-2^{n-m+1}\left(\begin{array}{c}2^{m-2}\\2\end{array}\right)\\
&& \ \ \ \ =2^{n+m-4}
\end{eqnarray*}

To illustrate the construction of $w^{(n)}$, the following example is given.

 Suppose that $n=5,m=3, j=5$, and  $Left(u^{(n)})=\{1000\ 1000\ 0000\ 0000\}$. Then
$Left(w^{(n)})$ can  be

$\{1000\ 1100\ 0100\ 0000\}$,

$\{1100\ 1100\ 0000\ 0000\}$,

$\{1000\ 1000\ 0100\ 0100\}$,

$\{1100\ 1000\ 0000\ 0100\}$,

$\{1000\ 1001\ 0001\ 0000\}$,

$\{1001\ 1001\ 0000\ 0000\}$,

$\{1000\ 1000\ 0001\ 0001\}$,

$\{1001\ 1000\ 0000\ 0001\}$.

Case A.2): There are only 2 nonzero elements with distance  $2^{n-1}$ among 4 nonzero elements of $u^{(n)}$.
First select 3 nonzero elements among 4 nonzero elements of $Left(w^{(n)})$, second select 2 nonzero elements among these 3 nonzero elements and each of these 2 nonzero elements can be put in $Left(u^{(n)})$ or $Right(u^{(n)})$. So the number of these $u^{(n)}$
 can be given by
$$2^{n+2m+j-10}\left(\begin{array}{c}4\\3\end{array}\right)\left(\begin{array}{c}3\\2\end{array}\right)\times 2^2=3\times 2^{n+2m+j-6}$$

Similar to the first case, it is easy to verify that each of these $u^{(n)}$ is counted $\frac{2^{n-1}}{2^{n-m+1}}=2^{m-2}$ times repeatedly.

Suppose that $u^{(n)}$ is a binary sequence with $W_H(u^{(n)})=4$,
and there are only 2 nonzero elements with distance  $2^{n-1}$ among 4 nonzero elements of $u^{(n)}$.
Then the absolute number of these $u^{(n)}$ can be given by
\begin{eqnarray*}&&
F3=3\times2^{n+2m+j-6}/2^{m-2}\\
&& \ \ \ \ \ =3\times2^{n+m+j-4}
\end{eqnarray*}

So, if  the 4 nonzero elements of $u^{(n)}$ are part of  the 8 nonzero elements of a $w^{(n)}$,
then there exist a binary sequence $v^{(n)}$ with $W_H(v^{(n)})=4$, such that $L(u^{(n)}+v^{(n)})=2^{n-1}-(2^{n-m}+2^{n-j})$.
So the 4-error linear complexity of $u^{(n)}+s^{(n)}$ is less than $2^{n-1}-(2^{n-m}+2^{n-j})$,
and the number of these $u^{(n)}$ can be given by
\begin{eqnarray*}&&
F4=2^{n+2m+j-10}\left(\begin{array}{c}8\\4\end{array}\right)-(2^{2(m-2)}-1)F1\\
&& \ \ \ \ \ \ \ \ \ -(2^{m+j-5}-1)F2-(2^{m-2}-1)F3\\
&& \ \ \ \ \ =2^{n+2m+j-6}+2^{n+m-4}+2^{n+j-4}+3\times2^{n+m+j-4}
\end{eqnarray*}

Case B): Consider $u^{(n)}$ that there does not exist any $v^{(n)}_1$, such that $L(u^{(n)}+v^{(n)}_1)=2^{n-1}-(2^{n-m}+2^{n-j})$, but  there  exists $v^{(n)}$, such that $L(u^{(n)}+v^{(n)})<2^{n-1}-(2^{n-m}+2^{n-j})$.
Let $t^{(n)}=s^{(n)}+u^{(n)}+v^{(n)}$, then $L(t^{(n)})=2^{n-1}-(2^{n-m}+2^{n-j})$, and $t^{(n)}+v^{(n)}=s^{(n)}+u^{(n)}$.

Case B.1): Investigate the sequence $w^{(n)}$ with $L(w^{(n)})=2^{n-1}-(2^{n-m}+2^{n-k}), m<k<j$ and $W_H(w^{(n)})=8$.
By Lemma 3.8, the number of these $w^{(n)}$ can be given by $2^{n+2m+k-10}$.

Suppose that $u^{(n)}$ is a binary sequence with $W_H(u^{(n)})=4$, and  the 4 nonzero elements are part of  the 8 nonzero elements of $w^{(n)}$. Then there are totally $2^{n+2m+k-10}\left(\begin{array}{c}8\\4\end{array}\right)$ sequences $u^{(n)}$, some of which are counted repeatedly.
And we will investigate these $u^{(n)}$ by 3 subcases.

Case B.1.1):  $Left(u^{(n)})=Right(u^{(n)})$.

By Case A.1, if there exist 2 nonzero elements whose distance is $2^{n-m}(1+2a)$,
then the number of these $u^{(n)}$ can be given by
$$F5=2^{n+m-4}$$

There exists a $v^{(n)}$, such that $L(u^{(n)}+v^{(n)})=2^{n-1}-(2^{n-m}+2^{n-j})$.
So there is no need to consider these $u^{(n)}$.

By Case A.1, if there exist 2 nonzero elements whose distance is $2^{n-k}(1+2a)$,
then the number of these $u^{(n)}$ can be given by
$$F6=2^{n+k-4}$$

And there exist {\scriptsize $$\left(\frac{2^{n-1}}{2^{n-m+1}}\right)^2+\left(\frac{2^{n-1}}{2^{n-m+1}}-1\right)^2
+\left(\frac{2^{n-1}}{2^{n-m+1}}-1\right)\times2=2^{2m-3}-1$$}  distinct binary sequences $v^{(n)}_i,1\le i\le2^{2m-3}-1,$
with $W_H(v^{(n)}_i)=4$, such that $L(u^{(n)}+v^{(n)}_i)\le 2^{n-1}-(2^{n-m}+2^{n-k})$.

To be specific, there exist $2^{2(m-2)}$ distinct binary sequences $v^{(n)}$,  such that $L(u^{(n)}+v^{(n)})= 2^{n-1}-(2^{n-m}+2^{n-k})$;
 there exist $(2^{m-2}-1)^2$ distinct binary sequences $v^{(n)}$,  such that $L(u^{(n)}+v^{(n)})= 2^{n-1}-(2^{n-m}+2^{n-k})$ or $2^{n-1}-2^{n-i}$, $2\le i<m$; there exist $2(2^{m-2}-1)$ distinct binary sequences $v^{(n)}$,  such that $W_H(u^{(n)}+v^{(n)})=4$.

To illustrate the construction of $v^{(n)}$, the following example is given.

 Suppose that $n=5,m=3,k=4$, and  $Left(u^{(n)})=\{1010\ 0000\ 0000\ 0000\}$. Then
$Left(u^{(n)}+v^{(n)})$ can  be

$\{1010\ 1010\ 0000\ 0000\}$

$\{1010\ 1000\ 0000\ 0010\}$

 $\{1010\ 0010\ 0000\ 1000\}$

  $\{1010\ 0000\ 0000\ 1010\}$

  $\{1010\ 0000\  1010\ 0000\}$

  $\{1000\ 0000\ 1000\ 0000\}$

    $\{0010\ 0000\  0010\ 0000\}$

Case B.1.2): There are only 2 nonzero elements whose distance is $2^{n-1}$.
By Case A.2), the number of these $u^{(n)}$ can be given by
$$F7=3\times2^{n+m+k-4}$$

There exist $\frac{2^{n-1}}{2^{n-m+1}}+(\frac{2^{n-1}}{2^{n-m+1}}-1)
=2^{m-1}-1$ distinct binary sequences $v^{(n)}_i,1\le i\le2^{m-1}-1$, with $W_H(v^{(n)}_i)=4$,
such that $L(u^{(n)}+v^{(n)}_i)\le 2^{n-1}-(2^{n-m}+2^{n-k})$.

To be specific, there exist $2^{m-2}$ distinct binary sequences $v^{(n)}$,  such that $L(u^{(n)}+v^{(n)})= 2^{n-1}-(2^{n-m}+2^{n-k})$;
 there exist $2^{m-2}-1$ distinct binary sequences $v^{(n)}$,   such that $W_H(u^{(n)}+v^{(n)})=4$.

To illustrate the construction of $v^{(n)}$, the following example is given.

 Suppose that $n=5,m=3,k=4$, and  $u^{(n)}=\{1010\ 1000\ 0000\ 0000\ 1000\ 0000\ 0000\ 0000\}$. Then
$Left(u^{(n)}+v^{(n)})$ can  be

$\{1010\ 1010\ 0000\ 0000\}$

$\{1010\ 1000\ 0000\ 0010\}$

  $\{1000\ 0000\ 1000\ 0000\}$


If  $v^{(n)}=\{0010\ 1000\ 1000\ 0000\ 0000\ 0000\ 1000\ 0000\}$, then $Left(w^{(n)})$ can  be

$\{0010\ 1010\ 1000\ 0000\}$ or

$\{0010\ 1000\ 1000\ 0010\}$

Case B.1.3): If there do not exist 2 nonzero elements whose distance is $2^{n-1}$, then the number of these $u^{(n)}$ can be given by \begin{eqnarray*}&&
F8=2^{n+2m+k-10}\times[\left(\begin{array}{c}4\\0\end{array}\right)+\left(\begin{array}{c}4\\1\end{array}\right)\\
&&\ \ \ \ \ \ \ \ \  +\left(\begin{array}{c}4\\2\end{array}\right)+\left(\begin{array}{c}4\\3\end{array}\right)+\left(\begin{array}{c}4\\4\end{array}\right)]\\
&& \ \ \ \ \ =2^{n+2m+k-6}
\end{eqnarray*}

There exists a binary sequence $v^{(n)}$, with $W_H(v^{(n)})=4$,
such that $L(u^{(n)}+v^{(n)})=2^{n-1}-(2^{n-m}+2^{n-k})<2^{n-1}-(2^{n-m}+2^{n-j}), m<k<j$.

Case B.2): Consider sequence $u^{(n)}$ that there does not exist any binary sequence $v^{(n)}$,
such that $L(u^{(n)}+v^{(n)})=2^{n-1}-(2^{n-m}+2^{n-k}), m<k<j$.

Let us divide one period of $u^{(n)}$ into $2^{n-m+1}$ subsequences
of form
$$\{u_a,u_{a+2^{n-m+1}}, u_{a+2^{n-m+2}},\cdots,
u_{a+(2^{m-1}-1)\times2^{n-m+1}}\}$$
where $0\le a<2^{n-m+1}.$

Case B.2.1): Suppose that $u^{(n)}$ is a $2^n$-periodic binary sequence with $W_H(u^{(n)})=2$,
and all 2 nonzero elements are in one  subsequence. Then the number of these $u^{(n)}$ can be given by
$$F9=2^{n-m+1}\left(\begin{array}{c}2^{m-1}\\2\end{array}\right)$$

Of these $u^{(n)}$, there are
$$F10=2^{n-1}$$
sequences, in which the distance of the 2 nonzero elements is $2^{n-1}$.
There exist $$\frac{2^{n-1}}{2\times2^{n-2}}+\frac{2^{n-1}}{2\times2^{n-3}}+\cdots+\frac{2^{n-1}}{2\times2^{n-m+1}}=2^{m-2}-1$$ distinct binary sequences $v^{(n)}_i,1\le i\le2^{m-2}-1$, with $W_H(v^{(n)}_i)=2$,
such that $L(u^{(n)}+v^{(n)}_i)=2^{n-1}-2^{n-r}, 1<r<m$.

So, if the 2 nonzero elements of $u^{(n)}$ are in one  subsequence, and the distance of the 2 nonzero elements is not $2^{n-1}$,
then the number of these $u^{(n)}$ can be given by
$$F11=F9-F10$$

There exists a binary sequence $v^{(n)}$ with $W_H(v^{(n)})=2$, such that $L(u^{(n)}+v^{(n)})=2^{n-1}-2^{n-r}, 1<r<m$.

Case B.2.2):
Suppose that $u^{(n)}$ is a $2^n$-periodic binary sequence with $W_H(u^{(n)})=4$,
and only 3 nonzero elements are in one subsequence. Then there exit at least one pair of nonzero elements with distance $2^{n-1}(1+2a), 1<i<m, a\ge0$. There is no such situation in Case B.1).

So  the number of these $u^{(n)}$ can be given by
$$F12=\left(\begin{array}{c}2^{n-m+1}\\2\end{array}\right)\times\left(\begin{array}{c}2\\1\end{array}\right)\times\left(\begin{array}{c}2^{m-1}\\3\end{array}\right)\times2^{m-1}$$

Of these $u^{(n)}$, there are
$$F13=\left(\begin{array}{c}2^{n-m+1}\\2\end{array}\right)\times\left(\begin{array}{c}2\\1\end{array}\right)\times2^{m-2}\times(2^{m-1}-2)\times2^{m-1}$$
sequences, in which the distance of the 2 nonzero elements is $2^{n-1}$.
It is easy to verify that there exists a binary sequence $v^{(n)}$ with $W_H(v^{(n)})=2$,
such that $L(u^{(n)}+v^{(n)})=2^{n-1}-2^{n-r}, 1<r<m.$

Let $t^{(n)}=s^{(n)}+u^{(n)}+v^{(n)}$, then $L(t^{(n)})=2^{n-1}-(2^{n-m}+2^{n-j})$, and $t^{(n)}+v^{(n)}=s^{(n)}+u^{(n)}$, thus we only need to count $t^{(n)}+v^{(n)}$.

So, if only 3 nonzero elements of $u^{(n)}$ are in one subsequence,
and there do not exist 2 nonzero elements whose distance is $2^{n-1}$,
then the number of these $u^{(n)}$ can be given by
$$F14=F12-F13$$

Of these $u^{(n)}$, there exist 3 distinct binary sequences $v^{(n)}_i,1\le i\le3$, with $W_H(v^{(n)}_i)=4$,
such that  $W_H(u^{(n)}+v^{(n)}_i)=4$ and  $L(u^{(n)}+v^{(n)}_i)<2^{n-1}-(2^{n-m}+2^{n-j})$.

Case B.2.3): Suppose that $u^{(n)}$ is a $2^n$-periodic binary sequence with $W_H(u^{(n)})=4$, and 2 nonzero elements
are in one subsequence, and the other 2 nonzero elements are in a different one. Then the number of these $u^{(n)}$ can be given by
\begin{eqnarray*}&&
F15=\left(\begin{array}{c}2^{n-m+1}\\2\end{array}\right)\times\left(\begin{array}{c}2^{m-1}\\2\end{array}\right)^2
\end{eqnarray*}

Of these $u^{(n)}$, if there exist 2 pair of nonzero elements with distance  $2^{n-1}$, then the distance of 2 nonzero elements from  different subsequences is $2^{n-i}(1+2a), m\le i\le n, a\ge 0$. If the 4 nonzero elements are in $w^{(n)}$, and $L(w^{(n)})=2^{n-1}-(2^{n-m}+2^{n-k}), m<k\le j$, then the number of those $u^{(n)}$ is $F2+\sum\limits_{k=m+1}^j 2^{n+k-4}$, which is  discussed in Case A.1) and Case B.1.1).
So the number of the remaining  sequences $u^{(n)}$ can be given by
$$F16=F15-F2-\sum\limits_{k=m+1}^j 2^{n+k-4}$$

Of these $u^{(n)}$, if there exist only 2 nonzero elements whose distance is $2^{n-1}$, then the distance of other   2 nonzero elements is
$2^{n-i}(1+2a), 1< i< m, a\ge 0$. Thus $u^{(n)}$ can not be in Case B.1).

The number of these $u^{(n)}$ can be given by
$$F17=\left(\begin{array}{c}2^{n-m+1}\\2\end{array}\right)\times\left(\begin{array}{c}2\\1\end{array}\right)\times2^{m-2}\times[\left(\begin{array}{c}2^{m-1}\\2\end{array}\right)-2^{m-2}]$$

There exist $(\frac{2^{n-1}}{2^{m+1}}-1)\times2+1=2^{m-1}-1$ distinct binary sequences $v^{(n)}_i,1\le i\le2^{m-1}-1$, with $W_H(v^{(n)}_i)=4$,
such that $L(u^{(n)}+v_i^{(n)})<2^{n-1}-(2^{n-m}+2^{n-j})$.

To illustrate the construction of $v^{(n)}$, the following example is given.

 Suppose that $n=5,m=3$, and  $u^{(n)}=\{1010\ 0000\ 0010\ 0000\ 1000\ 0000\ 0000\ 0000\}$. Then
$Left(u^{(n)}+v^{(n)})$ can  be

$\{1000\ 0000\ 1000\ 0000\}$

$\{1010\ 0000\ 1010\ 0000\}$

$\{0010\ 0000\  0010\ 0000\}$

If the distance of 2 nonzero elements in each subsequence is not $2^{n-1}$,
then the number of these $u^{(n)}$ can be given by
\begin{eqnarray*}&&
F18=\left(\begin{array}{c}2^{n-m+1}\\2\end{array}\right)\times[\left(\begin{array}{c}2^{m-1}\\2\end{array}\right)-2^{m-2}]^2
\end{eqnarray*}

There exist 3 distinct binary sequences $v^{(n)}_i,1\le i\le3$, with $W_H(v^{(n)}_i)=4$,
such that  $L(u^{(n)}+v^{(n)}_i)<2^{n-1}-(2^{n-m}+2^{n-j})$.

So, there are
$$F19=F16-F17-F18$$
sequences, in which the distance of 2 nonzero elements in each subsequence is $2^{n-1}$,
and   the distance of 2 nonzero elements from  different subsequences   is $2^{n-k}(1+2a), j<k\le n$.
There exist $(\frac{2^{n-1}}{2^{n-m+1}}-1)^2+(\frac{2^{n-1}}{2^{n-m+1}}-1)\times2=2^{2m-4}-1$
distinct binary sequences $v^{(n)}_i,1\le i\le2^{2m-4}-1$, with $W_H(v^{(n)}_i)=4$,
such that  $L(u^{(n)}+v^{(n)}_i)<2^{n-1}-(2^{n-m}+2^{n-j})$.

To illustrate the construction of $v^{(n)}$, the following example is given.

 Suppose that $n=5,m=3,j=4, k=5$, and  $Left(u^{(n)})=\{1100\ 0000\ 0000\ 0000\}$. Then
$Left(u^{(n)}+v^{(n)})$ can  be

$\{1100\ 0000\ 1100\ 0000\}$

$\{1000\ 0000\ 1000\ 0000\}$

$\{0100\ 0000\  0100\ 0000\}$

Case B.2.4): Suppose that $u^{(n)}$ is a binary sequence with $W_H(u^{(n)})=4$
and 2 nonzero elements of $u^{(n)}$ are in one subsequence and the other 2 nonzero elements are in 2 other distinct subsequences.
Then the number of these $u^{(n)}$ can be given by
\begin{eqnarray*}&&
F20=\left(\begin{array}{c}2^{n-m+1}\\3\end{array}\right)\times\left(\begin{array}{c}3\\1\end{array}\right)\times\left(\begin{array}{c}2^{m-1}\\2\end{array}\right)\times(2^{m-1})^2
\end{eqnarray*}

If the distance of 2  nonzero elements of the same subsequence is $2^{n-1}$, then we need to remove some sequences $u^{(n)}$  discussed in Case A.2) and Case B.1.2). The number of those $u^{(n)}$ is $3\times2^{n+m+k-4}$,
and there  exists a binary sequence $v^{(n)}$ with $W_H(v^{(n)})=4$,
such that $L(u^{(n)}+v^{(n)})=2^{n-1}-(2^{n-m}+2^{n-k}), m<k\le j$.

So the number of the remaining sequences $u^{(n)}$ can be given by
$$F21=F20-3\sum \limits_{k=m+1}^j 2^{n+m+k-4}$$

Of these $u^{(n)}$, if there do not exist any 2 nonzero elements whose distance is $2^{n-1}$, then the number of these $u^{(n)}$ can be given by
{\small\begin{eqnarray*}
F22=\left(\begin{array}{c}2^{n-m+1}\\3\end{array}\right)\left(\begin{array}{c}3\\1\end{array}\right)\times(\left(\begin{array}{c}2^{m-1}\\2\end{array}\right)-2^{m-2})\times(2^{m-1})^2
\end{eqnarray*}}

There exists one binary sequences $v^{(n)}$, with $W_H(v^{(n)})=4$, such that $L(u^{(n)}+v^{(n)})=2^{n-1}-2^{n-r}, 1<r<m$.

So, if there exist 2 nonzero elements of $u^{(n)}$ with distance  $2^{n-1}$, and
there does not
exist a binary sequence $v^{(n)}$ with $W_H(v^{(n)})=4$, such that
$L(u^{(n)}+v^{(n)})=2^{n-1}-(2^{n-m}+2^{n-k}), m<k\le j$,
then the number of these $u^{(n)}$ can be given by
$$F23=F21-F22$$

There exist $\frac{2^{n-1}}{2^{n-m+1}}-1=2^{m-2}-1$ distinct binary
sequences $v^{(n)}_i,1\le i\le2^{m-2}-1$, with $W_H(v^{(n)}_i)=4$,
such that $L(u^{(n)}+v^{(n)}_i)=2^{n-1}-2^{n-r}, 1<r<m$.

Case B.2.5): Suppose that $u^{(n)}$ is a binary sequence with
$W_H(u^{(n)})=4$ and all 4 nonzero elements of $u^{(n)}$ are in one
subsequence. Then the number of these $u^{(n)}$ can be given by
$$F24=2^{n-m+1}\times\left(\begin{array}{c}2^{m-1}\\4\end{array}\right)$$

Of these $u^{(n)}$, there are
$$F25=2^{n-m+1}\times\left(\begin{array}{c}2^{m-2}\\2\end{array}\right)$$
sequences, in which
there exist  nonzero elements $z_1$,$z_2, z_3$ and $z_4$, such that the distance of  $z_1$  and $z_2$, and the distance of  $z_3$  and $z_4$ are all
 $2^{n-1}$.

There exists a binary sequence $v^{(n)}$ with $W_H(v^{(n)})=0$, such that $L(u^{(n)}+v^{(n)})=2^{n-1}-2^{n-r}, 1<r<m$.
Let $t^{(n)}=s^{(n)}+u^{(n)}+v^{(n)}=s^{(n)}+u^{(n)}$.
Then $L(t^{(n)})=L(s^{(n)})=2^{n-1}-(2^{n-m}+2^{n-j})$.

If there are only 2 nonzero elements whose distance is $2^{n-1}$,
then the number of these $u^{(n)}$ can be given by
$$F26=2^{n-m+1}\times2^{m-2}\times[\left(\begin{array}{c}2^{m-1}-2\\2\end{array}\right)-(2^{m-2}-1)]$$

There exist 2 binary sequences $v^{(n)}$ with $W_H(v^{(n)})=2$,
such that $L(u^{(n)}+v^{(n)})=2^{n-1}-2^{n-r}, 1<r<m$.

So, if there do not exist 2 nonzero elements whose distance is $2^{n-1}$,
then the number of these $u^{(n)}$ can be given by
$$F27=F24-F25-F26$$

There exist $\left(\begin{array}{c}4\\2\end{array}\right)+1=7$ distinct binary sequences $v^{(n)}_i,1\le i\le7$, with $W_H(v^{(n)}_i)=4$,
such that  $L(u^{(n)}+v^{(n)}_i)<2^{n-1}-(2^{n-m}+2^{n-j})$.

By Lemma 3.2, the number of $2^{n}$-periodic binary sequences with given linear complexity $2^{n-2}-(2^{n-m}+2^{n-j})$ is
$2^{2^{n-2}-(2^{n-m}+2^{n-j})-1}$. It follows that

{\scriptsize
\begin{eqnarray*}&&N_4(2^{n-1}-(2^{n-m}+2^{n-j}))\\
&=&[1+\left(\begin{array}{c}2^n\\2\end{array}\right)+\left(\begin{array}{c}2^n\\4\end{array}\right)-F4\\
&&-\sum\limits_{k=m+1}^{j-1}(\frac{2^{2m-3}-1}{2^{2m-3}}F6+\frac{2^{m-1}-1}{2^{m-1}}F7+F8/2)\\
&&-\frac{2^{m-2}-1}{2^{m-2}}F10-F11/2-F13-\frac{3}{4}F14-\frac{2^{m-1}-1}{2^{m-1}}F17\\
&&-\frac{3}{4}F18-\frac{2^{2m-4}-1}{2^{2m-4}}F19-F22/2-\frac{2^{m-2}-1}{2^{m-2}}F23\\
&&-F25-F26-\frac{7}{8}F27]\times2^{2^{n-1}-(2^{n-m}+2^{n-j})-1}
\end{eqnarray*}}
\end{proof}\

We define $N_4(2^{n-1}-(2^{n-m}+2^{n-j}))=p(n,m,j)\times2^{2^{n-1}-(2^{n-m}+2^{n-j})-1}$\\

Next we   investigate the category of
$2^{n-1}-(2^{n-m}+2^{n-j})+x$.

\noindent {\bf Lemma  3.10}  Let $N_4(2^{n-1}-(2^{n-m}+2^{n-j})+x)$ be the number of $2^n$-periodic
binary sequences with linear complexity less than $2^n$ and given 4-error linear complexity
$2^{n-1}-(2^{n-m}+2^{n-j})+x, n>5, 2<m<j<n-1, 1\le x<2^{n-j-1}$.
Then
{\scriptsize
\begin{eqnarray*}&&N_4(2^{n-1}-(2^{n-m}+2^{n-j})+x)\\
&=&[1+\left(\begin{array}{c}2^n\\2\end{array}\right)+\left(\begin{array}{c}2^n\\4\end{array}\right)-\frac{2^{m+j-4}-1}{2^{m+j-4}}G1\\
&&\ \ -\sum\limits_{k=m+1}^j(\frac{2^{2m-3}-1}{2^{2m-3}}G2+\frac{2^{m-1}-1}{2^{m-1}}G3+G4/2)\\
&&\ \ -\frac{2^{m-2}-1}{2^{m-2}}G6-G7/2-G9-\frac{3}{4}G10-\frac{2^{m-1}-1}{2^{m-1}}G13\\
&&\ \ -\frac{3}{4}G14-\frac{2^{2m-4}-1}{2^{2m-4}}G15-G18/2-\frac{2^{m-2}-1}{2^{m-2}}G19\\
&&\ \ -G21-G22-\frac{7}{8}G23]2^{2^{n-1}-(2^{n-m}+2^{n-j})+x-1}
\end{eqnarray*}}
where $G1, G2,\cdots,G23$ are defined in  the following proof.

\begin{proof}\
Suppose that $s^{(n)}$ is a binary sequence  with linear complexity $2^{n-1}-(2^{n-m}+2^{n-j})+x$,
$u^{(n)}$ is a binary sequence with $W_H(u^{(n)})=2$ or 4.

Next we will investigate the $u^{(n)}$ by Case A) and Case B).

Case A): Suppose that $w^{(n)}$ is a $2^n$-periodic sequence with linear complexity $2^{n-1}-(2^{n-m}+2^{n-k}), m<k\le j$ and $W_H(w^{(n)})=8$,
By Lemma 3.8, the number of these $w^{(n)}$ can be given by $2^{n+2m+k-10}$.

Suppose that $u^{(n)}$ is a $2^n$-periodic binary sequence with $W_H(u^{(n)})=4$, and  the 4 nonzero elements are part of  the 8 nonzero elements of a $w^{(n)}$.
It is easy to verify that there exists a binary sequence $v^{(n)}$ with $W_H(v^{(n)})=4$, such that $L(u^{(n)}+v^{(n)})=2^{n-1}-(2^{n-m}+2^{n-k})<2^{n-1}-(2^{n-m}+2^{n-j})+x$, $m<k\le j$.

The discussion here is similar to Case B.1) in the proof of Lemma 3.9.

We will investigate these $u^{(n)}$ by the following 3 cases.

Case A.1): $Left(u^{(n)})=Right(u^{(n)})$. Based on the  proof of
Lemma 3.9,  the number of these $u^{(n)}$ can be given by
$$G0=2^{n+m-4}+2^{n+k-4}$$

Of these $u^{(n)}$, if there exist 2 nonzero elements in $Left(u^{(n)})$ whose distance is $2^{n-m}(1+2a), a\ge0$,
and the number of these $u^{(n)}$ can be given by
$$G1=2^{n+m-4}$$

Suppose that $W_H(v^{(n)})=4$ and $L(u^{(n)}+v^{(n)})<2^{n-1}-(2^{n-m}+2^{n-j})+x$.
We will investigate the number of these $v^{(n)}$ by the following 2 cases.

Case A.1.1): $L(u^{(n)}+v^{(n)})=2^{n-1}-(2^{n-m}+2^{n-k})$, $m<k\le j$. By Case A.1) of Lemma  3.9, the the number of these $v^{(n)}$ is $2^{m+k-5}$.

Case A.1.2): Let us divide $Left(u^{(n)})$ into $2^{n-m}$ subsequences. From the subsequence which contains the 2 nonzero elements of $Left(u^{(n)})$, we first select one location as one   nonzero element of $Left(v^{(n)})$. The number of locations of  another   nonzero element of $Left(v^{(n)})$ is $\frac{2^{n-1}}{2^{n-m+1}}$, such that the distance of 2 nonzero elements of $Left(v^{(n)})$ is $2^{n-m}(1+2a), a\ge0$.

As we select every $v^{(n)}$ twice and one $v^{(n)}$ is the same as $u^{(n)}$, thus the number of these $v^{(n)}$ is $$\frac{1}{2}(\frac{2^{n-1}}{2^{n-m}}\times\frac{2^{n-1}}{2^{n-m+1}})-1=2^{2(m-2)}-1,$$ such that $L(u^{(n)}+v^{(n)})<2^{n-1}-(2^{n-m}+2^{n-j})$.

Therefore, there exist
\begin{eqnarray*}&&\sum\limits_{k=m+1}^j2^{m+k-5}+2^{2(m-2)}-1=2^{m+j-4}-1\end{eqnarray*}
  distinct binary sequences $v^{(n)}_i,1\le i\le2^{m+j-4}-1$, with $W_H(v^{(n)}_i)=4$, such that $L(u^{(n)}+v^{(n)}_i)<2^{n-1}-(2^{n-m}+2^{n-j})+x$.
Let $t^{(n)}_i=s^{(n)}+u^{(n)}+v^{(n)}_i$,
then $L(s^{(n)})=L(t^{(n)}_i)=2^{n-1}-(2^{n-m}+2^{n-j})+x$ and $s^{(n)}+u^{(n)}=t^{(n)}_i+v^{(n)}_i$.

 The following example is given to illustrate the construction of $v^{(n)}$.

 Suppose that $n=6,m=3,j=4$, and  $Left(u^{(n)})=\{1000\ 0000\  1000\ 0000\ 0000\ 0000\ 0000\ 0000\}$. Then
$Left(u^{(n)}+v^{(n)})$ can  be

$\{1000\ 1000\ 1000\ 1000\ 0000\ 0000\ 0000\ 0000\}$

$\{1000\ 1000\ 1000\ 0000\ 0000\ 0000\ 0000\ 1000 \}$

$\{1000\ 0000\ 1000\ 1000\ 0000\ 1000\ 0000\ 0000\}$

$\{1000\ 0000\ 1000\ 0000\ 0000\ 1000\ 0000\ 1000\}$

$\{1000\  0000\ 0000\ 0000\ 1000\ 0000\ 0000\ 0000\}$

$\{0000\ 0000\ 1000\  0000\ 0000\ 0000\ 1000\  0000\}$

$\{1000\ 0000\ 1000\ 0000\ 1000\ 0000\ 1000\ 0000\}$

Based on the  proof of
Lemma 3.9, if there exist 2 nonzero elements in $Left(u^{(n)})$ whose
distance is $2^{n-k}(1+2a)$, then the number of these $u^{(n)}$ can
be given by
$$G2=G0-G1=2^{n+k-4}$$

Suppose that $W_H(v^{(n)})=4$ and $L(u^{(n)}+v^{(n)})<2^{n-1}-(2^{n-m}+2^{n-k})+x$.
We will investigate the number of these $v^{(n)}$ by the following 2 cases.

Case A.1.3): $L(u^{(n)}+v^{(n)})=2^{n-1}-(2^{n-m}+2^{n-k})$. By Case A.1) of Lemma  3.9, the the number of these $v^{(n)}$ is $2^{2(m-2)}$.

Case A.1.4):  From the locations  which have distance $2^{n-m}(2a), a\ge0$ with  the first nonzero element of $Left(u^{(n)})$, we first select one location as one   nonzero element of $Left(v^{(n)})$. Then from the locations  which have distance $2^{n-m}(2b), b\ge0$ with  the second nonzero element of $Left(u^{(n)})$, we  select one location as another   nonzero element of $Left(v^{(n)})$.

As  one $v^{(n)}$ is the same as $u^{(n)}$, thus the number of these $v^{(n)}$ is $$(\frac{2^{n-1}}{2^{n-m+1}})^2-1=2^{2(m-2)}-1,$$ such that $L(u^{(n)}+v^{(n)})<2^{n-1}-(2^{n-m}+2^{n-k})$.

Therefore, there exist

 \begin{eqnarray*}2^{2(m-2)}+2^{2(m-2)}-1=2^{2m-3}-1\end{eqnarray*} distinct binary sequences
$v^{(n)}_i,1\le i\le2^{2m-3}-1,$ with $W_H(v^{(n)}_i)=4$, such that
$L(u^{(n)}+v^{(n)}_i)<2^{n-1}-(2^{n-m}+2^{n-j})+x$.

Case A.2): There are only 2 nonzero elements whose distance is
$2^{n-1}$. By the proof of Lemma 3.9,  the number of these $u^{(n)}$
can be given by
$$G3=3\times2^{n+m+k-4}$$

There exist
$$\frac{2^{n-1}}{2^{n-m+1}}+(\frac{2^{n-1}}{2^{n-m+1}}-1)
=2^{m-1}-1$$ distinct binary sequences $v^{(n)}_i,1\le
i\le2^{m-1}-1$, with $W_H(v^{(n)}_i)=4$, such that
$L(u^{(n)}+v^{(n)}_i)<2^{n-1}-(2^{n-m}+2^{n-k})+x$.
There are $\frac{2^{n-1}}{2^{n-m+1}}-1$ distinct binary sequences $v^{(n)}$, such that $W_H(u^{(n)}+v^{(n)})=4$.

 The following example is given to illustrate the construction of $v^{(n)}$.

 Suppose that $n=6,m=3,k=4$, and  $Left(u^{(n)})=\{1000\ 1000\  1000\ 0000\ 0000\ 0000\ 0000\ 0000\}$. Then
$Left(u^{(n)}+v^{(n)})$ can  be

$\{1000\ 1000\ 1000\ 1000\ 0000\ 0000\ 0000\ 0000\}$

$\{1000\ 1000\ 1000\ 0000\ 0000\ 0000\ 0000\ 1000 \}$

$\{1000\  0000\ 0000\ 0000\ 1000\ 0000\ 0000\ 0000\}$

Case A.3): If there do not exist 2 nonzero elements whose distance
is $2^{n-1}$, then the number of these $u^{(n)}$ can be given by
\begin{eqnarray*}&&
G4=2^{n+2m+k-10}\times[\left(\begin{array}{c}4\\0\end{array}\right)+\left(\begin{array}{c}4\\1\end{array}\right)\\
&& \ \ \ \ \ \ \ \ +\left(\begin{array}{c}4\\2\end{array}\right)+\left(\begin{array}{c}4\\3\end{array}\right) +\left(\begin{array}{c}4\\4\end{array}\right)]\\
&& \ \ \ \ \ =2^{n+2m+k-6}
\end{eqnarray*}

There  exists a binary sequence $v^{(n)}$, with $W_H(v^{(n)})=4$,
such that $L(u^{(n)}+v^{(n)})=2^{n-1}-(2^{n-m}+2^{n-k})$.

Case B): Consider $u^{(n)}$ that there does not exist any $v^{(n)}_1$, such that $L(u^{(n)}+v^{(n)}_1)=2^{n-1}-(2^{n-m}+2^{n-k})$, but  there  exists $v^{(n)}$, such that $L(u^{(n)}+v^{(n)})<2^{n-1}-(2^{n-m}+2^{n-k})$.
Let $t^{(n)}=s^{(n)}+u^{(n)}+v^{(n)}$, then $L(t^{(n)})=2^{n-1}-(2^{n-m}+2^{n-j})+x$, and $t^{(n)}+v^{(n)}=s^{(n)}+u^{(n)}$.

The discussion here is similar to Case B.2) in the proof of Lemma 3.9.

Let us divide one period of $u^{(n)}$ into $2^{n-m+1}$ subsequences
of form
$$\{u_a,u_{a+2^{n-m+1}}, u_{a+2^{n-m+2}},\cdots,
u_{a+(2^{m-1}-1)\times2^{n-m+1}}\}$$
where $0\le a<2^{n-m+1}.$

Case B.1): Suppose that $u^{(n)}$ is a $2^n$-periodic binary
sequence with $W_H(u^{(n)})=2$, and both 2 nonzero elements are in
one subsequence. Then the number of these $u^{(n)}$ can be given by
$$G5=2^{n-m+1}\left(\begin{array}{c}2^{m-1}\\2\end{array}\right)$$

Of these $u^{(n)}$, there are
$$G6=2^{n-1}$$
sequences, in which the distance of the 2 nonzero elements is $2^{n-1}$.
There  exist $2^{m-2}-1$ distinct binary sequences $v^{(n)}_i,1\le i\le2^{m-2}-1$, with $W_H(v^{(n)}_i)=2$,
such that $L(u^{(n)}+v^{(n)}_i)=2^{n-1}-2^{n-r}, 1<r<m$.

So, if the 2 nonzero elements of $u^{(n)}$ are in one  subsequence,
and the distance of the 2 nonzero elements is not $2^{n-1}$,
then the number of these $u^{(n)}$ can be given by
$$G7=G5-G6$$

There  exists a binary sequence $v^{(n)}$ with $W_H(v^{(n)})=2$,
such that $L(u^{(n)}+v^{(n)})=2^{n-1}-2^{n-r}, 1<r<m$.

Case B.2): Suppose that $u^{(n)}$ is a $2^n$-periodic binary
sequence with $W_H(u^{(n)})=4$, and there are 3 nonzero elements  in
one  subsequence.
Then there exit at least one pair of nonzero elements with distance $2^{n-1}(1+2a), 1<i<m, a\ge0$. There is no such situation in Case A).

The number of these $u^{(n)}$ can be given by
$$G8=\left(\begin{array}{c}2^{n-m+1}\\2\end{array}\right)\times\left(\begin{array}{c}2\\1\end{array}\right)\times\left(\begin{array}{c}2^{m-1}\\3\end{array}\right)\times2^{m-1}$$

Of these $u^{(n)}$, there are
$$G9=\left(\begin{array}{c}2^{n-m+1}\\2\end{array}\right)\times\left(\begin{array}{c}2\\1\end{array}\right)\times2^{m-2}\times(2^{m-1}-2)\times2^{m-1}$$
sequences, in which there exist 2 nonzero elements with distance
$2^{n-1}$. It is easy to verify that there exists  a
binary sequence $v^{(n)}$ with $W_H(v^{(n)})=2$, such that
$L(u^{(n)}+v^{(n)})=2^{n-1}-2^{n-r}, 1<r<m$.

If there are 3 nonzero elements of $u^{(n)}$  in one  subsequence, and there do not exist 2 nonzero elements with distance  $2^{n-1}$,
then the number of these $u^{(n)}$ can be given by
$$G10=G8-G9$$

There exist 3 distinct binary sequences $v^{(n)}_i,1\le i\le3$, with
$W_H(v^{(n)}_i)=4$, such that $W_H(u^{(n)}+v^{(n)}_i)=4$ and
$L(u^{(n)}+v^{(n)}_i)<2^{n-1}-(2^{n-m}+2^{n-j})+x$.

Case B.3): Suppose that $u^{(n)}$ is a $2^n$-periodic binary
sequence with $W_H(u^{(n)})=4$, and 2 nonzero elements are in one
subsequence, and the other 2 nonzero elements are in a different
one. Then the number of these $u^{(n)}$ can be given by
\begin{eqnarray*}&&
G11=\left(\begin{array}{c}2^{n-m+1}\\2\end{array}\right)\times\left(\begin{array}{c}2^{m-1}\\2\end{array}\right)^2
\end{eqnarray*}

If there  exists a binary sequence $v^{(n)}$ with $W_H(v^{(n)})=4$,
such that $L(u^{(n)}+v^{(n)})=2^{n-1}-(2^{n-m}+2^{n-k}), m<k\le j$,
then the number of those $u^{(n)}$ is $G1+\sum\limits_{k=m+1}^j G2$.
So the number of the remaining  sequences $u^{(n)}$ can be given by
$$G12=G11-(G1+\sum\limits_{k=m+1}^j G2)$$

Of these $u^{(n)}$, if there exist only 2 nonzero elements whose distance is $2^{n-1}$ ,
then there exits one pair of nonzero elements with distance $2^{n-1}(1+2a), 1<i<m, a\ge0$. There is no such situation in Case A).

The number of these $u^{(n)}$ can be given by
$$G13=\left(\begin{array}{c}2^{n-m+1}\\2\end{array}\right)\times\left(\begin{array}{c}2\\1\end{array}\right)\times2^{m-2}\times[\left(\begin{array}{c}2^{m-1}\\2\end{array}\right)-2^{m-2}]$$

There  exist $(2^{m-2}-1)\times2+1=2^{m-1}-1$ distinct binary sequences $v^{(n)}_i,1\le i\le2^{m-1}-1$, with $W_H(v^{(n)}_i)=4$,
such that $L(u^{(n)}+v^{(n)}_i)=2^{n-1}-2^{n-r}$ or $2^{n-1}-(2^{n-r}+2^{n-k}), 1<r<m, m\le k\le n$.

 The  construction of $v^{(n)}$ is similar to that of B.2.3) in the proof of Lemma 3.9.

If the distance of 2 nonzero elements in each subsequence is not $2^{n-1}$,
then the number of these $u^{(n)}$ can be given by
\begin{eqnarray*}&&
G14=\left(\begin{array}{c}2^{n-m+1}\\2\end{array}\right)\times[\left(\begin{array}{c}2^{m-1}\\2\end{array}\right)-2^{m-2}]^2
\end{eqnarray*}

There  exist 3 distinct binary sequences $v^{(n)}_i,1\le i\le3$,
with $W_H(v^{(n)}_i)=4$, such that
$L(u^{(n)}+v^{(n)}_i)=2^{n-1}-2^{n-r}$ or $2^{n-1}-(2^{n-r}+2^{n-k}), 1<r<m, m\le k\le n$.

So, there are
$$G15=G12-G13-G14$$

sequences, in which the distance of 2 nonzero elements in each
subsequence is all $2^{n-1}$, and the distance of 2 nonzero elements from   different
subsequences is $2^{n-k}(1+2a), j<k\le n$. There  exist
$(\frac{2^{n-1}}{2^{n-m+1}}-1)^2+(\frac{2^{n-1}}{2^{n-m+1}}-1)\times2=2^{2m-4}-1$
distinct binary sequences $v^{(n)}_i,1\le i\le2^{2m-4}-1$, with
$W_H(v^{(n)}_i)=4$, such that
$L(u^{(n)}+v^{(n)}_i)=2^{n-1}-2^{n-r}$ or $2^{n-1}-(2^{n-r}+2^{n-k}), 1<r<m, j< k\le n$.

 The  construction of $v^{(n)}$ is similar to that of B.2.3) in the proof of Lemma 3.9.

Case B.4): Suppose that $u^{(n)}$ is a binary sequence with
$W_H(u^{(n)})=4$ and 2 nonzero elements of $u^{(n)}$ are in a
subsequence and the other 2 nonzero elements are in 2 other distinct
subsequences. Then the number of these $u^{(n)}$ can be given by
$$G16=\left(\begin{array}{c}2^{n-m+1}\\3\end{array}\right)\times\left(\begin{array}{c}3\\1\end{array}\right)\times\left(\begin{array}{c}2^{m-1}\\2\end{array}\right)\times(2^{m-1})^2$$

If  the distance of 2 nonzero elements  in a
subsequence is $2^{n-1}$ and
there
exists a binary sequence $v^{(n)}$ with $W_H(v^{(n)})=4$, such that
$L(u^{(n)}+v^{(n)})=2^{n-1}-(2^{n-m}+2^{n-k}), m<k\le j$.
By Case A.2),  the number of those $u^{(n)}$ is
$3\times(\sum\limits_{k=m+1}^j 2^{n+m+k-4})$.

So the number of the remaining sequences $u^{(n)}$ can be given by
$$G17=G16-3\times(\sum\limits_{k=m+1}^j 2^{n+m+k-4})$$

If there do not exist  2 nonzero elements with distance  $2^{n-1}$,
then the number of these $u^{(n)}$ can be given by
\begin{eqnarray*}&&G18\\
&=&\left(\begin{array}{c}2^{n-m+1}\\3\end{array}\right)\times\left(\begin{array}{c}3\\1\end{array}\right)[\left(\begin{array}{c}2^{m-1}\\2\end{array}\right)-2^{m-2}](2^{m-1})^2
\end{eqnarray*}

There  exists a binary sequences $v^{(n)}$, with $W_H(v^{(n)})=4$,
such that $L(u^{(n)}+v^{(n)})=2^{n-1}-2^{n-r}, 1<r<m$.

So, if  the distance of 2 nonzero elements  in a
subsequence is $2^{n-1}$ and
there
does not exist a binary sequence $v^{(n)}$ with $W_H(v^{(n)})=4$, such that
$L(u^{(n)}+v^{(n)})=2^{n-1}-(2^{n-m}+2^{n-k}), m<k\le j$,
then the number of these $u^{(n)}$ can be given by
$$G19=G17-G18$$

There exist $\frac{2^{n-1}}{2^{n-m+1}}-1=2^{m-2}-1$ distinct binary
sequences $v^{(n)}_i,1\le i\le2^{m-2}-1$, with $W_H(v^{(n)}_i)=4$,
such that $L(u^{(n)}+v^{(n)}_i)=2^{n-1}-2^{n-r}, 1<r<m$.

Case B.5):  Suppose that $u^{(n)}$ is a binary sequence with
$W_H(u^{(n)})=4$ and all 4 nonzero elements of $u^{(n)}$ are in one
subsequence. Then the number of these $u^{(n)}$ can be given by
$$G20=2^{n-m+1}\times\left(\begin{array}{c}2^{m-1}\\4\end{array}\right)$$

Of these $u^{(n)}$, there are
$$G21=2^{n-m+1}\times\left(\begin{array}{c}2^{m-2}\\2\end{array}\right)$$
sequences, in which there exist  nonzero elements $z_1$,$z_2, z_3$
and $z_4$, such that the distance of  $z_1$  and $z_2$, and the
distance of  $z_3$  and $z_4$ are all
 $2^{n-1}$.
There  exists a binary sequence $v^{(n)}$ with $W_H(v^{(n)})=0$, and $L(u^{(n)}+v^{(n)})=2^{n-1}-2^{n-r}, 1<r<m$.
Let $t^{(n)}=s^{(n)}+u^{(n)}+v^{(n)}$.
Then $L(t^{(n)})=L(s^{(n)})=2^{n-1}-(2^{n-m}+2^{n-j})+x$ and $t^{(n)}=s^{(n)}+u^{(n)}$.

If there are only 2 nonzero elements whose distance is $2^{n-1}$,
then the number of these $u^{(n)}$ can be given by
$$G22=2^{n-m+1}\times2^{m-2}\times[\left(\begin{array}{c}2^{m-1}-2\\2\end{array}\right)-(2^{m-2}-1)]$$

It is easy to verify that there 2 binary sequences
$v^{(n)}$ with $W_H(v^{(n)})=2$, such that
$L(u^{(n)}+v^{(n)})=2^{n-1}-2^{n-r}, 1<r<m$.

So, if there do not exist 2 nonzero elements whose distance is $2^{n-1}$,
then the number of these $u^{(n)}$ can be given by
$$G23=G20-G21-G22$$

There  exist $\left(\begin{array}{c}4\\2\end{array}\right)+1=7$
distinct binary sequences $v^{(n)}_i,1\le i\le7$, with
$W_H(v^{(n)}_i)=4$, such that
$L(u^{(n)}+v^{(n)}_i)<2^{n-1}-(2^{n-m}+2^{n-j})+x$. One of these $v^{(n)}$ is got by $Left(v^{(n)})=Right(u^{(n)})$ and $Right(v^{(n)})=Left(u^{(n)})$.

By Lemma 3.2, the number of $2^{n}$-periodic binary sequences with given linear complexity $2^{n-2}-(2^{n-m}+2^{n-j})+x$ is
$2^{2^{n-2}-(2^{n-m}+2^{n-j})+x-1}$. It follows that

{\scriptsize
\begin{eqnarray*}&&N_4(2^{n-1}-(2^{n-m}+2^{n-j})+x)\\
&=&[1+\left(\begin{array}{c}2^n\\2\end{array}\right)+\left(\begin{array}{c}2^n\\4\end{array}\right)-\frac{2^{m+j-4}-1}{2^{m+j-4}}G1\\
&&\ \ -\sum\limits_{k=m+1}^j(\frac{2^{2m-3}-1}{2^{2m-3}}G2+\frac{2^{m-1}-1}{2^{m-1}}G3+G4/2)\\
&&\ \ -\frac{2^{m-2}-1}{2^{m-2}}G6-G7/2-G9-\frac{3}{4}G10-\frac{2^{m-1}-1}{2^{m-1}}G13\\
&&\ \ -\frac{3}{4}G14-\frac{2^{2m-4}-1}{2^{2m-4}}G15-G18/2-\frac{2^{m-2}-1}{2^{m-2}}G19\\
&&\ \ -G21-G22-\frac{7}{8}G23]2^{2^{n-1}-(2^{n-m}+2^{n-j})+x-1}
\end{eqnarray*}}
\end{proof}\

We define $N_4(2^{n-1}-(2^{n-m}+2^{n-j})+x)=q(n,m,j)\times2^{2^{n-1}-(2^{n-m}+2^{n-j})+x-1}$\\

\noindent {\bf Lemma  3.11}  Let $L(r,c)=2^n-2^r+c, 4\le r\le n,
1\le c\le 2^{r-3}-1$, and $N_4(L)$ be the number of
$2^n$-periodic binary sequences with  linear complexity less than
$2^n$ and given 4-error linear complexity $L$. Then {\small
$$N_4(L)=\left\{\begin{array}{l}
1+\left(\begin{array}{c}2^n\\2\end{array}\right)+\left(\begin{array}{c}2^n\\4\end{array}\right),\
\ \ \ \ \ \ \ L=0\ \   \\
2^{L-1}(1+\left(\begin{array}{c}2^r\\2\end{array}\right)+\left(\begin{array}{c}2^r\\4\end{array}\right)),
 \ L=L(r,c)
\end{array}\right.$$}
\begin{proof}\
Suppose that $s$ is a binary sequence with first period
$s^{(n)}=\{s_0,s_1,s_2,\cdots, s_{2^n-1}\}$, and $L(s^{(n)})=2^n-2^r+c$. By
Games-Chan algorithm, $Left(s^{(t)})\ne Right(s^{(t)}), r+1\le t\le n$,
 where
$s^{(t)}=\varphi_{t+1}\cdots\varphi_{n}(s^{(n)})$.
\

First consider the  case of $W_H(s^{(n)})=0$. There is only one
binary sequence of this kind.
\

Consider the  case of $W_H(s^{(n)})=2$. There is 2 nonzero bits in
$\{s_0,s_1,\cdots, s_{2^n-1}\}$, thus there are
$\left(\begin{array}{c}2^n\\2\end{array}\right)$ binary sequences of
this kind.

\
Consider the  case of $W_H(s^{(n)})=4$. There is 4 nonzero bits in
$\{s_0,s_1,\cdots, s_{2^n-1}\}$, thus there are
$\left(\begin{array}{c}2^n\\4\end{array}\right)$ binary sequences of
this kind.
\

So $N_4(0)=1+\left(\begin{array}{c}2^n\\2\end{array}\right)+\left(\begin{array}{c}2^n\\4\end{array}\right)$.
\

Consider $L(r,c)=2^n-2^r+c$, $4\le r\le n, 1\le c\le 2^{r-3}-1$.
Suppose that $s^{(n)}$ is a binary sequence with
$L(s^{(n)})=L(r,c)$. Note that
$L(r,c)=2^n-2^r+c=2^{n-1}+\cdots+2^r+c$.  By Games-Chan algorithm,
$Left(s^{(r)})= Right(s^{(r)})$, and $L(s^{(r)})=c$.

It is known that the number of  binary sequences $t^{(r)}$ with
$W_H(t^{(r)})=0$, 2 or 4 is
$1+\left(\begin{array}{c}2^r\\2\end{array}\right)+\left(\begin{array}{c}2^r\\4\end{array}\right)$.
\

By Lemma 3.4, the 4-error linear complexity of $s^{(r)}+t^{(r)}$ is
$c$. \

By Lemma 3.2 and Lemma 3.3, the number of  binary sequences
$s^{(r)}+t^{(r)}$ is $2^{c-1}\times
(1+\left(\begin{array}{c}2^r\\2\end{array}\right)+\left(\begin{array}{c}2^r\\4\end{array}\right))$
\

By Lemma 3.1, there are $2^{2^{n-1}+\cdots+2^r}=2^{2^{n}-2^r}$
binary sequences $s^{(n)}+t^{(n)}$,  such that
$s^{(r)}+t^{(r)}=\varphi_{r+1}\cdots\varphi_{n}(s^{(n)}+t^{(n)})$,
$t^{(r)}=\varphi_{r+1}\cdots\varphi_{n}(t^{(n)})$ and
$W_H(t^{(n)})=W_H(t^{(r)})$.
\

Thus  the 4-error linear complexity of
$s^{(n)}+t^{(n)}$ is
$${2^{n-1}+\cdots+2^r}+L_4(s^{(r)}+t^{(r)})={2^{n}-2^r}+c=L(r,c).$$
\

Therefore, $N_4(L(r,c))=2^{2^{n}-2^r}\times 2^{c-1}\times
(1+\left(\begin{array}{c}2^r\\2\end{array}\right)+\left(\begin{array}{c}2^r\\4\end{array}\right))
=2^{L(r,c)-1}(1+\left(\begin{array}{c}2^r\\2\end{array}\right)+\left(\begin{array}{c}2^r\\4\end{array}\right))$.
\end{proof}\

Based on the results above, now we can have the proof of Theorem 3.1.
%
%

\begin{proof}\

By Lemma 3.11, we now only need to consider $2^{r-3}\le
c\le 2^{r-1}-1$.

By Lemma 3.1 and Lemma 3.5, $N_4(L(r,c))=2^{L(r,c)-1}f(r,m)$ for
$3\le r\le n,   c= 2^{r-2}-2^{r-m}, 2<m\le r$

By Lemma 3.1 and Lemma 3.6, $N_4(L(r,c))=2^{L(r,c)-1}g(r,m)$ for
$5\le r\le n,   c= 2^{r-2}-2^{r-m}+x,2<m<r-1,0<x<2^{r-m-1}$

By Lemma 3.1 and Lemma 3.7, $N_4(L(r,c))=2^{L(r,c)-1}h(r,m)$ for
$2\le r\le n,   c= 2^{r-1}-2^{r-m}, 2\le m\le r$

By Lemma 3.1 and Lemma 3.9, $N_4(L(r,c))=2^{L(r,c)-1}p(r,m,j)$ for
$4\le r\le n,   c= 2^{r-1}-(2^{r-m}+2^{r-j}),2<m<j\le r$

By Lemma 3.1 and Lemma 3.10, $N_4(L(r,c))=2^{L(r,c)-1}q(r,m,j)$ for
$6\le r\le n, c= 2^{r-1}-(2^{r-m}+2^{r-j})+x,2<m<j<r-1,0<x<2^{r-j-1}$

This completes the proof.
\end{proof}\

\

The following is an example to illustrate Theorem 3.1.

For $n=5$, the number of $2^n$-periodic binary sequences with linear
complexity less than $2^n$ is $2^{2^5-1}=2147483648$.

From Theorem 3.1, the numbers of $2^n$-periodic binary sequences
with  linear complexity less than $2^n$ and given  4-error linear
complexity for $n = 5$ are shown in Table 1, and these results are
also checked by computer.

\begin{center}
\begin{tabular}{ll}
Table 1. $N_4(L(r,c))$& by Theorem 3.1\\
\hline $L(r,c)$&  $N_4(L(r,c))$\  \\
\hline
 0 & 36457   \\
 1 & 36457   \\
 2 & 72914   \\
 3 & 145828   \\
 4 & 289416   \\
 5 & 581072   \\
 6 & 1144608   \\
 7 & 2236992   \\
 8 & 2293760   \\
 9 & 6837504   \\
 10 & 13210112   \\
 11 & 25031680   \\
 12 & 14876672   \\
 13 & 46845952   \\
 14 & 8978432   \\
 15 & 4587520  \\
 16 & 0   \\
 17 & 127205376   \\
 18 & 236060672   \\
 19 & 418643968   \\
 20 & 134217728   \\
 21 & 567279616   \\
 22 & 33554432   \\
 23 & 16777216   \\
 24 & 0   \\
 25 & 486539264   \\
 26 & 0   \\
 27 & 0   \\
 28 & 0   \\
 29 & 0   \\
 30 & 0   \\
 31 & 0  \\
\hline
\end{tabular}
\end{center}

The summation of numbers of the right column is  $2^{31}$.

 \section*{ Acknowledgment}
 The research was supported by
Anhui Natural Science Foundation(No.1208085MF106) and NSAF
(No. 10776077).

\end{document}